\input epsf
\scrollmode 
\overfullrule=0mm
\hsize=4.7truein \vsize=7.0truein \hoffset=.2truein 
\global\newcount\meqno \global\meqno=1                                   
\def\eqn#1#2{\xdef #1{(\the\meqno)}\global\advance\meqno by1 $$#2\eqno#1$$}  
\global\newcount\refno \global\refno=1 \newwrite\rfile                    
\def\ref#1#2{$^{\the\refno}$\nref#1{#2}}%
\def\nref#1#2{\xdef#1{$^{\the\refno}$}%
\ifnum\refno=1\immediate\openout\rfile=refs.tmp\fi%
\immediate\write\rfile{\noexpand\item{\the\refno .\ }#2}%
\global\advance\refno by1}                                                
   
\def\rsc{$^,$} 
\global\newcount\figno \global\figno=1    
\def\figure#1#2#3{\midinsert\centerline{\epsffile{#2}}
\xdef #1{\the\figno}\global\advance\figno by1
\smallskip\centerline{\vtop{\hsize 4.1in \noindent {\bf Figure #1.}~#3}}
\bigskip\endinsert} 
\def\nofigure#1#2#3{\midinsert\vskip#2 truecm
\xdef #1{\chapsym.\the\figno}\global\advance\figno by1
\hfil{{\bf #1}~#3}\hfil\endinsert}                                                                   
\font\bms=cmbsy10
\def\frac#1#2{{#1\over #2}}

\def\bx{{\bf x}}
\def\bq{{\bf q}}
\def\br{{\bf r}}

\def\bF{{\bf F}}

\def\bv{{\bf v}}

\def\rh{{\hat r}}
\def\Rh{{\hat R}}

\def\beps{{\bms \varepsilon}}

\def\dx{{d^d\bx\,}}
\def\dt{{dt\,}}
\def\dq{{d^d\bq \over (2\pi)^d}}
\def\dw{{d\omega \over 2\pi}}
\def\w{{\omega}}
\def\K{{\cal K}}
\def\del{\partial}    \def\frac#1#2{{#1\over#2}}
   \def\br{{\bf r}}
\def\para{\parallel} 
\def\hl{{r_\parallel}}
\def\htr{{r_\perp}}
\def\lal{\lambda_\para} \def\lat{\lambda_\perp} \def\lalt{\lambda_\times}
\def\Kl{{K_\para}} \def\Kt{{K_\perp}}
\def\fl{{f_\para}} \def\ft{{f_\perp}}
\def\Tl{{T_\para}} \def\Tt{{T_\perp}}
\def\zetal{{\zeta_\para}} \def\zetat{{\zeta_\perp}}
\def\zl{{z_\para}} \def\zt{{z_\perp}}
\centerline{\bf NONEQUILIBRIUM DYNAMICS OF INTERFACES AND LINES} 
\bigskip
\centerline{Mehran Kardar}
\smallskip
\centerline{\it Department of Physics}
\centerline{\it Massachusetts Institute of Technology}
\centerline{\it Cambridge, Massachusetts 02139, USA }
\bigskip
\noindent
{These notes are prepared for a set of lectures delivered at the
{\it The 4th CTP Workshop on Statistical Physics:
``Dynamics of Fluctuating Interfaces and Related Phenomena"}, 
at Seoul National University, Korea.
The lectures examine several problems related to non-equilibrium fluctuations of interfaces
and flux lines. The first two introduce the phenomenology of depinning, with particular
emphasis on interfaces and contact lines. The role of the anisotropy of the medium
in producing different universality classes is elucidated. The last two lectures focus
on the dynamics of lines, where transverse fluctuations are also important.
We shall demonstrate how various non-linearities appear in the dynamics of
driven flux lines. The universality classes of depinning, and also dynamic roughening,
are illustrated in the contexts of moving flux lines, advancing crack fronts, 
and drifting polymers.}
\vskip .15truein 
\noindent{\bf 1. Depinning of Interfaces}
\medskip
\noindent{\it 1.1 Introduction and Phenomenology}
\smallskip\noindent
Depinning is a non-equilibrium critical phenomenon involving an external force
and a pinning potential. When the force is weak the system is stationary,
trapped in a metastable state. Beyond a threshold force the (last) metastable   
state disappears and the system starts to move. 
A simple example is provided by a point mass on a rough table.
The mass is stationary until the external force $F$ exceeds that of static friction
$F_c$. Larger forces lead to an initial period of acceleration, before the motion
settles to a uniform velocity due to viscous forces. In the latter is proportional to
velocity, the ultimate velocity of the point close to threshold behaves as
$v\propto (F-F_c)$.

While there are many other macroscopic mechanical examples, 
our main interest comes from condensed matter systems
such as Charge Density Waves (CDWs)\ref
\rCDWs{H. Fukuyama and P. A. Lee, Phys. Rev. B {\bf 17}, 535 (1978);
P. A. Lee and T. M. Rice, Phys. Rev. B {\bf 19}, 3970 (1979).}, 
interfaces\ref
\rIntdepin{R. Bruinsma and G. Aeppli, Phys. Rev. Lett. {\bf 52}, 1547 (1984);
J. Koplik and H. Levine, Phys. Rev. B {\bf 32}, 280 (1985).}, 
and contact lines\ref
\rDeGennes{P.G.~de~Gennes, Rev. Mod. Phys. {\bf 57},
827 (1985).}.
In CDWs, the control parameter is the external voltage; a finite CDW current
appears only beyond a threshold applied voltage. Interfaces in porous
media, domain walls in random magnets, are stationary unless the applied
force (magnetic field) is sufficiently strong. A key feature of these
examples is that they involve the {\it collective} depinning of many degrees
of freedom that are elastically coupled. As such, these problems belong to
the realm of collective critical phenomena, characterized by universal
scaling laws. We shall introduce these laws and the corresponding 
exponents below for the depinning of a line (interface or contact line). 

\epsfysize=5cm\figure\CL{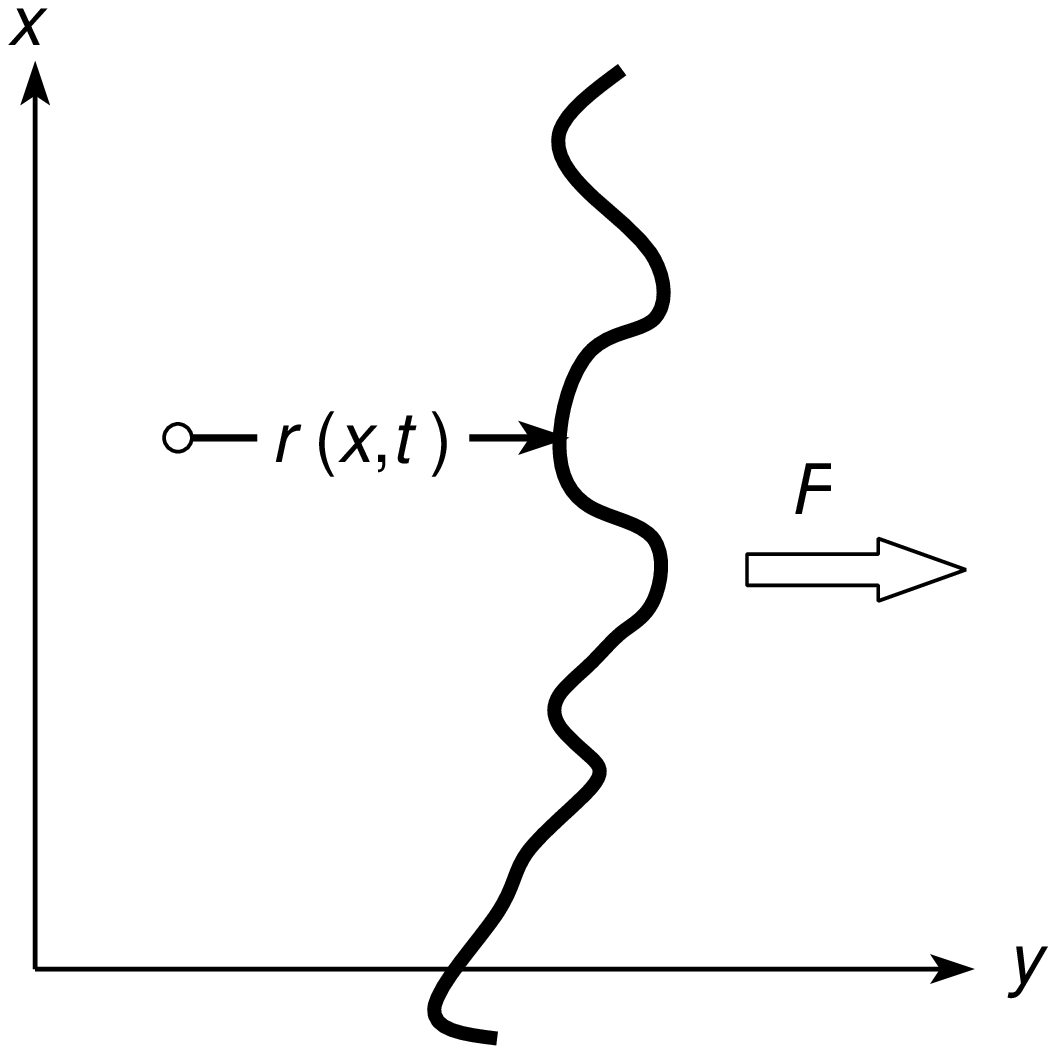}{Geometry of the line in two dimensions.}
Consider a line in two dimensions, oriented along the $x$ direction, and 
fluctuating along a perpendicular direction $r$. The configuration of the
line at time $t$ is described by the function $r(x,t)$. 
The function $r$  is assumed
to be single valued, thus excluding configurations with overhangs.  In
many cases\rIntdepin, where viscous forces dominate over inertia, the local velocity
of a point on the curve is given by
\eqn\eGen{{d{r(x,t)} \over dt}=F+f(x,r)+{\cal K}[r].}
The first term on the right hand side is a uniform applied force which is also the external control 
parameter. Fluctuations in the force due to randomness and impurities 
are represented by the second term. With the assumption that the medium
is on average translationally invariant, the average of $f$ can be set to zero. 
The final term in eq.\eGen describes the elastic forces between different parts of the
line. Short range interactions can be described by a gradient expansion; 
for example, a line tension leads to ${\cal K}[r(x)]=\nabla^2 r$ or
${\cal K}[r(q)]=-q^2 r(q)$ for the Fourier modes. The surface of a drop
of non--wetting liquid terminates at a {\it contact line} on a solid substrate\rDeGennes.
Deformations of the contact line are accompanied by distortions of the
liquid/gas surface. As shown by Joanny and de Gennes\ref
\rJoanny{J.F.~Joanny and P.G.~de~Gennes, J. Chem. Phys.
{\bf 81}, 552 (1984).}, 
the resulting energy and forces are {\it non--local}, described by ${\cal K}[r(q)]=-|q| r(q)$.

For the case of a surface in three dimensions deformations are described by
$r(x_1,x_2)$. More generally, we shall consider $r({\bf x})$, where ${\bf x}$ is a
$d-$dimensional vector. In a similar spirit, we shall generalize the coupling to 
${\cal K}[r(q)]=-|q|^\sigma r(q)$,
which interpolates between the above two cases as $\sigma$ changes
from one to two. Note, however, that the equation of motion need not originate
from variations of a Hamiltonian, and may include non-linear couplings which 
will be discussed later on.

\epsfysize=5cm\figure\fVvsF{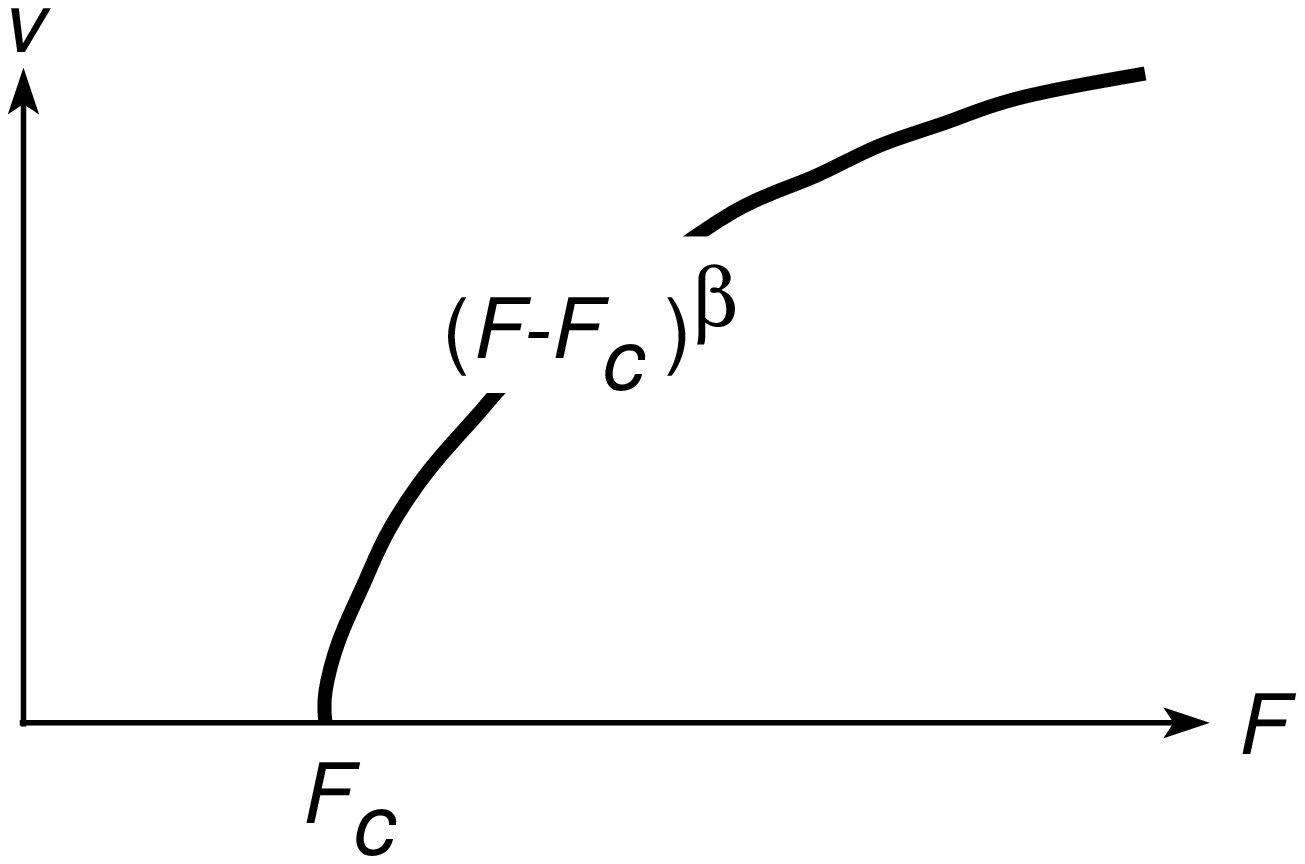}{Critical behavior of the velocity.}
When $F$ is small, the line is trapped in one of many metastable states in which
$\partial r/\partial t=0$ at all points. For $F$ larger than a threshold $F_c$, 
the line is depinned from the last metastable state, and moves with an average 
velocity $v$. On approaching the threshold from above, the velocity vanishes as
\eqn\eVel{v=A(F-F_c)^\beta,}
where $\beta$ is the {\it velocity exponent}, and $A$ is a
nonuniversal amplitude. A mean-field estimate for $\beta$ was obtained by
Fisher in the context of 
CDWs\ref \rDSF{D.S. Fisher, Phys. Rev. Lett. {\bf 50}, 1486 (1983).}. It
corresponds to the limit $\sigma=0$, where every point is coupled to all others,
and hence experiences a restoring force proportional to 
$\left\langle r({\bf x)} \right\rangle- r({\bf x})$. The resulting equation of motion,
$${dr({\bf x})\over dt}=\left\langle r({\bf x)} \right\rangle- r({\bf x})+F+f({\bf x},r({\bf x})),$$
has to be supplemented with the condition $\left\langle r({\bf x)} \right\rangle=vt$.
The self-consistent solution for the velocity indeed vanishes as $(F-F_c)^\beta$,
with an exponent that depends on the details of the random force. If $f({\bf x},r({\bf x}))$
varies smoothly with $r$, the exponent is $\beta=3/2$, while discontinuous jumps
in the force (like a saw--tooth) result in $\beta=1$. In fact the latter is a better starting
point for depinning in finite dimensions. This is because of the avalanches in
motion (discussed next), which lead to a discontinuous coarse grained force.

The motion just above threshold is not uniform,
composed of rapid jumps as large segments of the line depin from strong
pinning centers, superposed on the slower steady advance. These
jumps have a power law distribution in size, cutoff at a correlation length
$\xi$ which diverges at the transition as
\eqn\eXi{\xi\sim (F-F_c)^{-\nu}.}
The jumps are reminiscent of {\it avalanches} in other slowly driven systems.
In fact, the depinning can be approached from below $F_c$ by monotonically
increasing $F$ in small increments, each sufficient to cause a jump to the
next metastable state. The size and width of avalanches becomes invariant
on approaching $F_c$. For example,
\eqn\eAval{{\rm Prob(width\ of\ avalanche} > \ell) \approx{1\over \ell^\kappa}
\hat\rho(\ell/\xi_-),}
where the cutoff $\xi_-$ diverges as in Eq.\eXi.
The critical line is a {\it self--affine} fractal whose correlations satisfy 
the dynamic scaling from
\eqn\eHcorr{\langle\left[r(x,t)-r(x',t')\right]^2\rangle
=(x-x')^{2\zeta}g\left({{|t-t'|}\over{|x-x'|^z}}\right),}
defining  the {\it roughness} and {\it dynamic} exponents, $\zeta$ and $z$
respectively. (Angular brackets reflect averaging over all realizations of
the random force $f$.) The scaling function $g$
goes to a constant as its argument approaches 0; $\zeta$ is
the wandering exponent of an instantaneous line profile, and
$z$ relates the average lifetime of an avalanche to its size
by $\tau(\xi)\sim\xi^z$.

Although, the underlying issues of collective depinning for CDWs
and interfaces have been around for some time, only recently a systematic 
perturbative approach to the problem was developed. This functional
renormalization group (RG) approach to the dynamical equations of
motion was originally developed in the context of CDWs by Narayan and
Fisher\ref\rCDW{D.S.~Fisher, Phys. Rev. B {\bf 31}, 1396 (1985), 
O.~Narayan and D.S.~Fisher, Phys. Rev. B {\bf 46}, 11520 (1992).}\  (NF), 
and extended to interfaces by Nattermann et al\ref
\rNSTL{T.~Nattermann, S.~Stepanow, L.-H.~Tang, and
H.~Leschhorn, J. Phys. II France {\bf 2}, 1483 (1992).}. 
We shall provide
a brief outline of this approach starting from Eq.\eGen. Before embarking
on the details of the formalism, it is useful to point out some scaling relations
amongst the exponents which follow from underlying symmetries and
non-renormalization conditions.
 
\noindent{\bf 1.} As mentioned earlier, the motion of the line close to the
threshold is composed of jumps of segments of size $\xi$. Such jumps
move the interface forward by $\xi^\zeta$ over a time period $\xi^z$.
Thus the velocity behaves as,
\eqn\eVid{v\sim {\xi^\zeta \over \xi^z}\sim|F-F_c|^{\nu(z-\zeta)}\quad
\Longrightarrow \quad \beta=\nu(z-\zeta).}

\noindent{\bf 2.} If the elastic couplings are linear, the response of the line to a 
{\it static} perturbation $\varepsilon (x)$ is obtained simply by considering
\eqn\ernew{r_\varepsilon (x,t)=r(x,t)-{\cal K}^{-1}[\varepsilon (x)],}
where ${\cal K}^{-1}$ is the inverse kernel. Since, $r_\varepsilon $ satisfies 
Eq.\eGen\ subject to a force $F+\varepsilon (x)+f(x,r_\varepsilon )$, $r$ satisfies 
the same
equation with a force $F+f(x,r-{\cal K}^{-1}[\varepsilon (x)])$. As long as the
statistical properties of the stochastic force are not modified by the above
change in its argument, ${\partial \left\langle r \right\rangle}/\partial \varepsilon =0$, and
\eqn\eSres{\left\langle {\partial {r_\varepsilon (x)}\over \partial \varepsilon (x)} 
\right\rangle=-{\cal K}^{-1},\quad {\rm or}\quad 
\left\langle {\partial {r_\varepsilon (q)}\over \partial \varepsilon (q)} \right\rangle
={1 \over |q|^\sigma}.}
Since it controls the macroscopic response of the line, the kernel ${\cal K}$  
cannot change under RG scaling. From Eqs.\eHcorr\ and \eXi, we can 
read off the scaling of $r(x)$, and the force $\delta F$, which using the
above non-renormalization must be related by the exponent relation
\eqn\eFid{\zeta+{1\over\nu}=\sigma.}
Note that this identity depends on the statistical invariance of noise under the
transformation in Eq.\ernew. It is satisfied as long as the force correlations
$\left\langle f(x,r)f(x',r') \right\rangle$ only depend on $r-r'$. The identity
does not hold if these correlations also depend on the slope
$ {\partial r/ \partial x}$.

\noindent {\bf 3.} A scaling argument related to the Imry--Ma estimate of the
lower critical dimension of the random field Ising model, can be used to
estimate the roughness exponent\ref
\rIM{Y. Imry and S.-K. Ma, Phys. Rev. Lett. {\bf 35}, 1399 (1975).}. 
The elastic force on a segment of 
length $\xi$ scales as $\xi^{\zeta-\sigma}$. If fluctuations in force are
uncorrelated in space, they scale as $\xi^{-(\zeta+1)/2}$ over the area of an 
avalanche. Assuming that these two forces must be of the same order 
to initiate the avalanche leads to
\eqn\eIMl{\zeta={2\sigma-1\over 3}.}
This last argument is not as rigorous as the previous two. Nonetheless, 
all three exponent identities can be established within the RG framework.
Thus the only undetermined exponent is the dynamic one, $z$.

\medskip
\noindent{\it 1.2 Functional Renormalization Group}
\smallskip\noindent
A field theoretical description of the dynamics of Eq.\eGen\ can be developed
using the formalism of Martin, Siggia and Rose\ref\rMSR{P.C.~Martin, 
E.~Siggia, and H.~Rose, Phys. Rev. A {\bf 8}, 423 (1973).}\  (MSR): Generalizing
to a $d-$dimensional interface, an auxiliary field 
$\rh(\bx,t)$ is introduced to implement the equation of motion as a series of 
$\delta$--functions. Various dynamical response and correlation functions for 
the field $r(\bx,t)$ can then be generated from the functional,
\eqn\eZl{Z=\int{{\cal D} r(\bx,t){\cal D}\rh(\bx,t){\cal J}[r]\exp(S)},} 
where
\eqn\eSl{S=i\int{d^d\bx\,\dt\rh(\bx,t)\left\{\partial_t r-{\cal K}[r]
-F-f\left({\bf x},r({\bf x},t)\right)\right\}}.}
The Jacobian ${\cal J}[r]$ is introduced to ensure that  the $\delta$--functions 
integrate to unity. It does not generate any new relevant terms and will be
ignored henceforth.

The disorder-averaged generating functional $\overline Z$ can be evaluated 
by a saddle-point expansion
around a Mean-Field (MF) solution obtained by setting 
$\K_{MF}[r(\bx)]=vt-r(\bx).$  
This amounts to replacing interaction forces with Hookean springs connected
to the center of mass, which moves with a velocity $v$. 
The corresponding equation of motion is
\eqn\eMF{{dr_{MF}\over dt}=vt-r_{MF}(t)+f[r_{MF}(t)]+F_{MF}(v),}
where the relationship $F_{MF}(v)$ between the external force $F$ and average
velocity $v$ is determined from the consistency condition 
$\langle r_{MF}(t)\rangle=vt$.
The MF solution depends on the type of irregularity\rCDW: 
For smoothly varying random potentials, 
$\beta_{MF}=3/2$, whereas for cusped random potentials, $\beta_{MF}=1$.
Following the treatment of NF\rCDW$^,$\ref\rNF{O.~Narayan and D.~S.~Fisher, 
Phys. Rev. B {\bf 48}, 7030 (1993).},
we use the mean field solution for cusped potentials,
anticipating jumps with velocity of $O(1)$, in which case
$\beta_{MF}=1$. After rescaling and averaging over impurity 
configurations, we arrive at a generating functional whose 
low-frequency form is
\eqn\eZav{\eqalign{
\overline Z &= \int{\cal D}R(\bx,t){\cal D}\Rh(\bx,t)\exp(\tilde S), \cr 
\tilde S &= -\int \dx\dt\left[F-F_{MF}(v)\right] \hat R(\bx,t) \cr
  &\qquad-\int\dq\dw \hat R(-\bq,-\w)(-i\w\rho+|\bq|^\sigma)R(\bq,\w) \cr
  &\qquad+ {1 \over 2}\int\dx\dt dt'\, \hat R(\bx,t)\Rh(\bx,t')
 C\left[vt-vt'+R(\bx,t)-R(\bx,t')\right].\cr}}
In the above expressions, $R$ and $\Rh$ are coarse-grained
forms of $r-vt$ and $i\rh$, respectively. $F$ is adjusted to satisfy
the condition $\langle R\rangle=0$. The function $C(v\tau)$ is
initially the connected mean-field correlation function
$\langle(r_{MF}(t)r_{MF}(t+\tau)\rangle_c$. 

Ignoring the $R$-dependent terms in the argument of $C$, the 
action becomes Gaussian, and is invariant under a scale
transformation $x\to bx$, $t\to b^\sigma t$, $R\to b^{\sigma-d/2}R$,
$\hat R\to b^{-\sigma-d/2} \hat R$, $F\to b^{-d/2}F$, and
$v\to b^{-d/2}v$. Other terms in the action, of higher
order in $R$ and $\hat R$, that result from the expansion of $C$
[and other terms not explicitly shown in Eq.\eZav],
decay away at large length and time scales
if $d>d_c=2\sigma$. For $d>d_c$, the interface is smooth ($\zeta_0<0$) 
at long length scales, and the depinning exponents take the Gaussian 
values $z_0=\sigma$, $\nu_0=2/d$, $\beta_0=1$.

At $d=d_c$, the action $S$ has an infinite number of marginal 
terms that can be rearranged as a Taylor series for the
function $C\left[vt-vt'+R({\bf x},t)-R({\bf
x},t')\right]$, when $v\to 0$. The RG is carried out by 
integrating over
a momentum shell $\Lambda/b<|\bq|<\Lambda$ (we set the
cutoff wave vector to $\Lambda=1$
for simplicity) and all frequencies, followed by a scale
transformation $x\to bx$, $t\to b^zt$, $R\to b^\zeta R$, and
$\hat R\to b^{\theta-d} \hat R$, where $b=e^\ell$.
The resulting recursion relation for the linear part
in the effective action (to all orders in perturbation theory) is
\eqn\eFrr{{\partial(F-F_{MF})\over \partial\ell}=(z+\theta)(F-F_{MF})+{\rm constant},}
which immediately implies (with a suitable definition of $F_c$)
\eqn\eFcrr{{\partial(F-F_c)\over \partial\ell}= y_F(F-F_c),}
with the exponent identity
\eqn\eTid{y_F=z+\theta=1/\nu\quad.} 
The functional renormalization of $C(u)$ in $d=2\sigma-\epsilon$ 
interface dimensions, computed to one-loop order,
gives the recursion relation,
\eqn\eCrr{\eqalign{
{\partial C(u) \over \partial \ell}=[\epsilon &+ 2\theta
+2(z-\sigma)]C(u)+\zeta u{dC(u)\over du} \cr
 &- {S_d\over (2\pi)^d}{d\over du}\left\{\left[C(u)-C(0)\right]
{dC(u)\over du}\right\},\cr}}
where $S_d$ is the surface area of a unit sphere in $d$ dimensions.
NF showed that all higher order diagrams contribute to the
renormalization of $C$ as
total derivatives with respect to $u$, thus, integrating
Eq.\eCrr\
at the fixed-point solution $\partial C^*/\partial\ell = 0$,
together with Eqs.\eFid\ and \eTid,
gives $\zeta=\epsilon/3$ to all orders in $\epsilon$,
provided that $\int C^*\neq 0$. This gives Eq.\eIMl\ for a 
one-dimensional interface, as argued earlier. 
This is a consequence of the fact that $C(u)$ remains short-ranged
upon renormalization, implying the absence of anomalous
contributions to $\zeta$. 

The dynamical exponent $z$ is calculated through the renormalization
of $\rho$, the term proportional to $\Rh\partial_tR$, which yields
\eqn\eZexp{z=\sigma-2\epsilon/9+O(\epsilon^2),}
and using the exponent identity \eVid,
\eqn\eBexp{\beta=1-2\epsilon/9\sigma+O(\epsilon^2).}
Nattermann et. al.\rNSTL\ obtain the same 
results to $O(\epsilon)$ by directly averaging the MSR generating
function in Eq.\eZl, and expanding perturbatively around a rigidly
moving interface.

Numerical integration of Eq.\eGen\ for an elastic interface\ref\rDong{
M.~Dong, M.~C.~Marchetti, A.~A.~Middleton, and V.~Vinokur, Phys. Rev.
Lett. {\bf 70}, 662 (1993). The identification of the exponent $\zeta=1$ from
correlation function has been questioned by 
H. Leschhorn and L.-H. Tang, Phys. Rev. Lett. {\bf 70}, 2973 (1993). }\ 
$(\sigma=2)$ 
has yielded critical exponents $\zeta=0.97\pm0.05$ and $\nu=1.05\pm0.1$,
in agreement with the theoretical result $\zeta=\nu=1$. The velocity 
exponent $\beta=0.24\pm0.1$ is also consistent with the one-loop theoretical
result 1/3; however, a logarithmic dependence $v\sim1/\ln(F-F_c)$, which
corresponds to $\beta=0$, also describes the numerical data well.
In contrast, experiments and various discrete models of interface growth 
have resulted in scaling behaviors that differ from system to system. 
A number of different experiments 
on fluid invasion in porous media\ref\rIntexp{M.A. Rubio, C.A. Edwards, 
A. Dougherty, and J.P Gollub, Phys. Rev. Lett. {\bf 63}, 1685 (1989);
V.K. Horv\'ath, F. Family, and T. Vicsek, Phys. Rev. Lett. {\bf 67}, 
3207 (1991); S.~He, G.~L.~M.~K.~S.~Kahanda, and P.-Z. Wong, Phys. Rev.
Lett. {\bf 69}, 3731 (1992).}\ 
give roughness exponents of around 0.8,
while imbibition experiments\ref\rBul{S.~V.~Buldyrev, 
A.-L.~Barabasi, F.~Caserta, S.~Havlin, H.~E.~Stanley, and T.~Vicsek, 
Phys. Rev. A {\bf 45}, R8313 (1992).}$^,$\ref
\eFCA{F.~Family, K.~C.~B.~Chan,
and J.~G.~Amar, in {\it Surface Disordering: Growth, Roughening and
Phase Transitions}, Les Houches Series, Nova Science Publishers, 
New York (1992).}\ 
have resulted in $\zeta\approx0.6$.
A discrete model studied by Leschhorn\ref\rHeiko{H.~Leschhorn, 
Physica A {\bf 195}, 324 (1993).}, motivated by Eq.\eGen\ with $\sigma=2$, 
gives a roughness exponent of 1.25 at threshold.
Since the expansion leading to Eq.\eGen\
breaks down when $\zeta$ approaches one, it is not clear how to reconcile
the results of Leschhhorn's numerical work\rHeiko\ with the coarse-grained
description of the RG calculation, especially since any model with $\zeta>1$ 
cannot have a coarse grained description based on gradient expansions. 

\medskip
\noindent{\it 1.3 Anisotropy}
\smallskip\noindent
Amaral, Barabasi, and Stanley (ABS)\ref\rAmar{L.~A.~N.~Amaral, 
A.-L.~Barabasi, and  H.~E.~Stanley, Phys. Rev. Lett. {\bf 73}, 62 (1994).}\ 
recently pointed out 
that various models of interface depinning in 1+1 dimensions fall
into two distinct classes, depending on the tilt dependence of 
the interface velocity:

\noindent {\bf 1.} For models like the random field Ising Model\ref
\rRFIM{H.~Ji and M.~O.~Robbins, Phys. Rev. B {\bf 44}, 2538 (1991);
B.~Koiller, H.~Ji, and M.~O.~Robbins, Phys. Rev. B {\bf 46}, 5258 (1992).},
and some Solid On Solid models, the computed exponents are consistent
with the exponents given by the RG analysis. It has been suggested\rHeiko,
however, that the roughness exponent is systematically larger than
$\epsilon/3$, casting doubt on the exactness of the RG result.

\noindent {\bf 2.} A number of different models, based on 
directed percolation (DP)\ref\rTL{L.-H.~Tang and H.~Leschhorn,
Phys. Rev. A {\bf 45}, R8309 (1992).}$^,$\rBul\ give a
different roughness exponent, $\zeta\approx0.63$. In these
models, pinning sites are randomly distributed with a probability
$p$, which is linearly related to the force $F$. The interface
is stopped by the boundary of a DP cluster of pinning sites. The 
critical exponents at depinning can then be related to the
longitudinal and transverse correlation length exponents 
$\nu_\parallel\approx1.70$ and $\nu_\perp\approx1.07$ of DP.
In particular, $\zeta=\nu_\parallel/\nu_\perp\approx0.63$, and
$\beta=\nu_\parallel-\nu_\perp\approx0.63$, in
agreement with experiments. 

The main difference of these models can be understood in terms
of the dependence of the threshold force $F_c$ to the orientation.
To include the possible dependence of the line mobility on its slope,
$\partial_x r$, we can generalize the equation of motion to
\eqn\etiltline{\partial_t r=K\partial_x^2 r + \kappa \partial_x r +
{\lambda\over2} (\partial_x r)^2 + F + f(x,r).}
The isotropic depinning studied by RG corresponds to $\kappa=\lambda=0$.
The usual mechanisms for generating a non-zero $\lambda$ are of kinematic origin\ref
\rKPZ{M. Kardar, G. Parisi, and Y.-C. Zhang, Phys. Rev. Lett. 
{\bf 56}, 889 (1986).}\ 
($\lambda\propto v$) and can be shown to be 
irrelevant at the depinning threshold where the velocity $v$ 
goes to zero\rNF.  However, if $\lambda$ 
{\it is not} proportional to $v$ and stays finite at the transition,
it is a relevant operator and expected to modify the critical behavior.
As we shall argue below, anisotropy in the medium is a possible source
of the nonlinearity at the depinning transition.

A model flux line (FL) confined to move in a plane\rDong\rsc\ref
\rFeng{C.~Tang, S.~Feng, and L.~Golubovic, Phys. Rev. Lett. {\bf 72}, 1264 (1994).}\ 
provides an 
example where both mechanisms for the nonlinearity are present. 
Only the force normal to the FL is responsible for motion,
and is composed of three components: {\bf (1)} A term proportional to
curvature arising from the smoothening effects of line tension.
{\bf (2)} The Lorentz force due to a uniform current density perpendicular to
the plane acts in the normal direction and has a uniform magnitude
$F$ (per unit line length). {\bf (3)} A random force 
${\hat {\bf n}}\cdot {\bf f}$
due to impurities, where ${\hat {\bf n}}$ is the unit normal vector\rFeng. 
Equating viscous dissipation with the work done by the normal force leads
to the equation of motion 
\eqn\eFaL{
{\partial h\over \partial t}=\sqrt{1+s^2}\left[
{\partial_x^2 h\over (1+s^2)^{3/2}}+F+{f_h-sf_x \over \sqrt{1+s^2}}\right],}
where $h(x,t)$ denotes transverse displacement of the line and
$s\equiv\partial_x h$. The  nonlinearities generated
by $\sqrt{1+s^2}$ are kinematic in origin\rKPZ\ and irrelevant
as $v\to0$\rNF, as can be seen easily by taking
them to the left hand side of Eq.\eFaL. The shape of the pinned
FL is determined by the competition of the terms in the
square brackets. Although there is no explicit simple $s^2$ term in this
group, it will be generated if the system is {\it anisotropic}.

To illustrate the idea, let us take $f_h$ and $f_x$ to be independent
random fields with amplitudes $\Delta_h^{1/2}$ and $\Delta_x^{1/2}$
respectively; each correlated isotropically in space within
a distance $a$. For weak disorder, a deformation of order $a$ 
in the normal direction $\hat{\bf n}$ takes place over a distance
$L_c\gg a$ along the line. The total force due to curvature on this piece
of the line is of the order of $L_c(a/L_c^2)$, and the pinning force,
$[(L_c/a)(n_h^2\Delta_h+n_x^2\Delta_x)]^{1/2}$.
Equating the two forces\rIntdepin\ yields 
$L_c=a(n_h^2\Delta_h+n_x^2\Delta_x)^{-1/3}$ and 
an effective pinning strength per unit length,
$$F_0(s)=aL_c^{-2}=a^{-1}\Bigl({\Delta_h+s^2\Delta_x\over 1+s^2}\Bigr)^{2/3}.$$
The roughening by impurities thus reduces
the effective driving force
on the scale $L_c$ to $\tilde F(s)=F-F_0(s)$.
Therefore, even if initially $F$ is independent of $s$, such
a dependence is generated under coarse graining, 
{\it provided that the random force
is anisotropic}, i.e. $\Delta_h\ne\Delta_x$. 
An expansion of $\tilde F(s)$ around its maximum (which defines the
hard direction) yields an $s^2$ term which is positive and 
remains finite as $v\rightarrow 0$.

The above example indicates the origin of the two types of behavior for
$\lambda_{\rm eff}=v''(s=0)$ 
observed by ABS\rAmar: 
Kinematics produces a $\lambda_{\rm eff}$ 
proportional to $v$ which vanishes at the threshold;
anisotropy yields a nonvanishing (and diverging) $\lambda_{\rm eff}$
at the depinning transition.
An immediate consequence of the latter is that the
depinning threshold $F_c$ depends on the average orientation of the line.
While anisotropy may generate other local terms in the 
effective equation of motion, at a symmetry direction,
this term is the only relevant one in the RG sense, capable of
modifying the critical behavior for $d\leq 4$.
A one-loop RG of Eq.\etiltline\ with the
$\kappa=0$ was carried out by Stepanow\ref
\rStepanow{S. Stepanow, J. Phys. II France {\bf 5}, 11 (1995).}. 
He finds no stable fixed point for $2\leq d\leq 4$, but his numerical
integration of the one loop RG equations in $d=1$ yield
$\zeta\approx 0.8615$ and a dynamical exponent $z=1$.
Due to the absence of Galilean invariance,
there is also a renormalization of $\lambda$ which is related
to the diverging $\lambda_{\rm eff}$ observed in
Ref.\rAmar.
The nonperturbative nature of the fixed point precludes a
gauge of the reliability of these exponents.

Numerical simulations of Eq.\etiltline\ in $d=1$\ref
\rNumDP{Z. Chah\'ok, K. Honda, and T. Vicsek,
J. Phys. A {\bf 26}, L171 (1993); 
S. Galluccio and Y.-C. Zhang, Phys. Rev. E {\bf 51}, 1686 (1995);
H. Leschhorn, cond-mat 9605018.},
indicate that it shares the characteristics
of a class of lattice models\rTL\rsc\rBul\ where the external force  
is related to the density $p$ of  ``blocking sites'' by $F=1-p$.
When $p$ exceeds a critical value of $p_c$, blocking sites form
a directed percolating path which stops the interface.
For a given geometry, there is a direction along which the first
spanning path appears. This defines a {\it hard} direction
for depinning where the threshold force $F_c(s)$ reaches maximum.
Higher densities of blocking sites are needed to form a spanning
path away from this direction, resulting in a lower
threshold force $F_c(s)$ for a tilted interface.
Thus on a phenomenological level we believe that the nonlinear equation,
and directed percolation (DP) models of interface depinning
belong to the same universality class of {\it anisotropic depinning}.
This analogy may in fact be generalized
to higher dimensions, where the blocking path
is replaced by a directed blocking surface\ref
\rDhar{D. Dhar, M. Barma, and M.K. Phani, Phys. Rev. Lett.
{\bf 47}, 1238 (1981); D. Dhar, J. Phys. A {\bf 15}, 1859 (1982);
S.V. Buldyrev, S. Havlin, and H.E. Stanley,
Physica {\bf 200}, 200 (1993); and references therein.}.
Unfortunately, little is known analytically about the scaling properties
of such a surface at the percolation threshold.

As emphasized above, the hallmark of anisotropic depinning
is the dependence of the threshold force $F_c(s)$ on the 
slope $s$.  Above this threshold, we expect 
$v(F,s)$ to be an analytical function of $F$ and $s$.
In particular, for $F>F_c(0)$, there is a small $s$ expansion
$v(F,s)=v(F,s=0)+\lambda_{\rm eff}s^2/2+\cdots$.
On the other hand, we can associate a characteristic slope 
$\overline{s}=\xi_\perp/
\xi_\parallel\sim(\delta F)^{\nu (1-\zeta)}$, to DP clusters
where $\delta F=F-F_c(0)$, and $\nu $ is the correlation length
exponent. Scaling then suggests
\eqn\evelocity{
v(F,s)=(\delta F)^{\theta}g(s/\delta F^{\nu (1-\zeta)}),}
where  $\theta=\nu (z-\zeta)$. Matching Eq.\evelocity\ with 
the small $s$ expansion, we see that $\lambda_{\rm eff}$ diverges 
as $(\delta F)^{-\phi}$ (as defined by ABS\rAmar) with 
$\phi=2\nu (1-\zeta)-\theta=\nu (2-\zeta-z)$.
In $d=1$, the exponents $\nu $ and $\zeta$ are
related to the correlation length exponents $\nu_\parallel$ and $\nu_\perp$
of DP\rDhar\ via $\nu =\nu_\parallel\approx 1.73$ and 
$\zeta=\nu_\perp/\nu_\parallel\approx 0.63$,
while the dynamical exponent is $z=1$.
Scaling thus predicts $\phi\approx 0.63$, in agreement with
the numerical result of $0.64\pm 0.08$ in Ref.\rAmar.
Close to the line $F=F_c(0)$ (but at a finite $s$), the dependence
of $v$ on $\delta F$ drops out and we have
\eqn\evFc{
v(F_c,s)\propto |s|^{\theta/ \nu (1-\zeta)}.}
As $z=1$ in $d=1$, the above equation 
reduces to $v\propto |s|$, in agreement with Fig.~1 of Ref.\rAmar.
Since $v(F,s)=0$ at $F=F_c(s)$,
Eq.\evelocity\ suggests
\eqn\eFc{
F_c(s)-F_c(0) \propto -|s|^{1/\nu (1-\zeta)}.}
Note that Eqs. \evFc\ and \eFc\ are valid also in higher
dimensions, though values of the exponents quoted above
vary with $d$\rDhar. 

An interface tilted away from the hard direction not only
has a different depinning threshold, but also completely different
scaling behavior at its transition. This is because,
due to  the presence of an average interface gradient 
${\bf s}=\left\langle \nabla h \right\rangle$, the isotropy 
in the internal ${\bf x}$ space is lost. The equation of motion
for fluctuations, $h'({\bf x},t)=h({\bf x},t)-{\bf s}\cdot{\bf x}$, around the 
average interface position may thus include a non-zero $\kappa$ in
\etiltline.  The resulting depinning transition belongs
to yet a new universality class with {\it anisotropic}
response and correlation functions in directions parallel
and perpendicular to ${\bf s}$; i.e.
$$\eqalign{\left\langle [h({\bf x})-h({\bf x'}) ]^2\right\rangle 
=&|x_\parallel-x_\parallel'|^\zeta
{\cal F}\left({|{\bf x_t}-{\bf x'_t}|\over|x_\parallel-x_\parallel'|^\eta}
\right)\cr
\to&\cases{
|x_\parallel-x_\parallel'|^\zeta& for ${\bf x_t}-{\bf x'_t}=0$\cr
|{\bf x_t}-{\bf x'_t}|^{\zeta/\eta}& for $x_\parallel-x_\parallel'=0 $}, }$$
where $\eta$ is the {\it ansiotropy} exponent, and ${\bf x}_t$
denotes the $d-1$ directions transverse to ${\bf s}$.

A suggestive mapping allows us to determine the exponents for
depinning a tilted interface: Consider the response to a perturbation
in which all points along a $(d-1)$-dimensional cross section of
the interface at a fixed $x_\parallel$ are pushed up by a small amount.
This move decreases the slope of the
interface uphill but increases it downhill.
Since $F_c(s)$ decreases with increasing $s$,
at criticality the perturbation propagates only a finite
distance uphill but causes a downhill avalanche.
The disturbance front moves at a constant
velocity ($\delta x_\parallel\propto t$) and hence $z_\parallel=1$.
(Such chains of moving sites were indeed seen in
simulations of the $d=2$ model discussed below.)
Furthermore, the evolution of successive cross sections 
${\bf x}_t(x_\parallel)$ is expected to be the same as the
evolution in time of a $(d-1)$-dimensional interface! 
The latter is governed by the Kardar-Parisi-Zhang (KPZ) equation\rKPZ, 
whose scaling behavior has
been extensively studied. From this analogy we conclude,
\eqn\etdpKPZ{
\zeta(d)={\zeta_{\rm KPZ}(d-1)\over z_{\rm KPZ}(d-1)},\quad
\eta(d)={1\over z_{\rm KPZ}(d-1)}.}
In particular, the tilted interface with
$d=2$ maps to the growth problem in 1+1 dimensions where the exponents
are known exactly, yielding $\zeta(2)=1/3$ and $\eta(2)=2/3$.
This picture can be made more precise for a lattice model introduced below.
Details will be presented elsewhere.

To get the exponent $\beta$ for the vanishing of velocity
of the tilted interface, we note that since $z_\parallel=1$,
$v$ scales as the excess slope $\delta s=s-s_c(F)$.
The latter controls the density of the above moving fronts;
$s_c(F)$ is the slope of the critical interface at a given driving force $F$,
i.e., $F=F_c(s_c)$.
Away from the symmetry direction,
the function $F_c(s)$ has a non-vanishing derivative and hence
\eqn\evtilt{\delta F=F-F_c(s)=F_c(s_c)-F_c(s)\sim \delta s\sim v.}
We thus conclude that generically $\beta=1$ for tilted interfaces, 
independent of dimension.

To check the above predictions, we performed simulations of
the parallelized version of a previously studied percolation model of interface 
depinning\rTL.
A solid-on-solid (SOS) interface is described by a set of integer heights 
$\{h_{\bf i}\}$
where ${\bf i}$ is a group of $d$ integers. With each configuration is 
associated a random set of 
pinning forces  $\{\eta_{\bf i}\in[0,1)\}$. The heights are updated 
{\it in parallel} according to the following rules:
$h_{\bf i}$ is increased by one if (i)
$h_{\bf i}\leq h_{\bf j}-2$ for at least one
${\bf j}$ which is a nearest neighbor of ${\bf i}$, {\it or} (ii)
$\eta_{\bf i}<F$ for a pre-selected uniform force $F$. If  
$h_{\bf i}$ is increased, the associated random force $\eta_{\bf i}$ is also
updated, i.e. replaced by a new random number in the interval $[0,1)$. 
Otherwise, $h_{\bf i}$ and $\eta_{\bf i}$ are unchanged.
The simulation is started with initial conditions
$h_{\bf i}(t=0)={\rm Int}[s {\bf i}_x]$, and 
boundary conditions $h_{\bf i+L}={\rm Int}[s L]+h_{\bf i}$ are 
enforced throughout. 
The CPU time is greatly 
reduced by only keeping track of active sites.

The above model has a simple analogy to
a resistor-diode percolation problem\rDhar.
Condition (i) ensures that, once a site $({\bf i},h)$ is wet (i.e.,
on or behind the interface),
all neighboring columns of {\bf i} must be wet
up to height $h-1$. Thus there is always ``conduction'' from a site
at height $h$ to sites in the neighboring columns at height $h-1$.
This relation can be represented by diodes pointing diagonally downward.
Condition (ii) implies that ``conduction'' may also occur upward.
Hence a fraction $F$ of vertical bonds are turned into resistors
which allow for two-way conduction. Note that, due to the SOS
condition, vertical downward conduction is always possible.
For $F<F_c$, conducting sites connected to a point lead
at the origin, form a cone whose hull is the interface separating
wet and dry regions. The opening angle of the cone
increases with $F$, reaching $180^\circ$ at $F=F_c$,
beyond which percolation in the entire space
takes place, so that all sites are eventually wet.
If instead of a point, we start with a planar lead defining
the initial surface, the percolation threshold depends
on the surface orientation, with the highest threshold
for the untilted one.

Our simulations of lattices of $65 536$ sites in $d=1$ and of $512\times 512$
and $840\times 840$ sites in $d=2$ confirm the exponents for depinning
in the hard direction.
For a tilted surface in $d=1$ the roughness exponent determined
from the height-height correlation function is consistent with
the predicted value of $\zeta={1/ 2}$ and different from 
$\zeta\approx 0.63$ of the untilted one.
The dependence of the depinning threshold on slope is clearly seen
in the figure below, where the average velocity is plotted against the driving
force for $s=0$ (open) and $s=1/2$ (solid).
The $s=0$ data can be fitted to a power-law
$v\sim (F-F_c)^\theta$, where $F_c\approx 0.461$, $\beta=0.63\pm 0.04$
for $d=1$, and $F_c\approx 0.201$, $\beta=0.72\pm 0.04$ for $d=2$.
Data at $s={1/ 2}$ are consistent with Eq.\evtilt\
close to the threshold.

\epsfysize=5cm\figure\TKDa{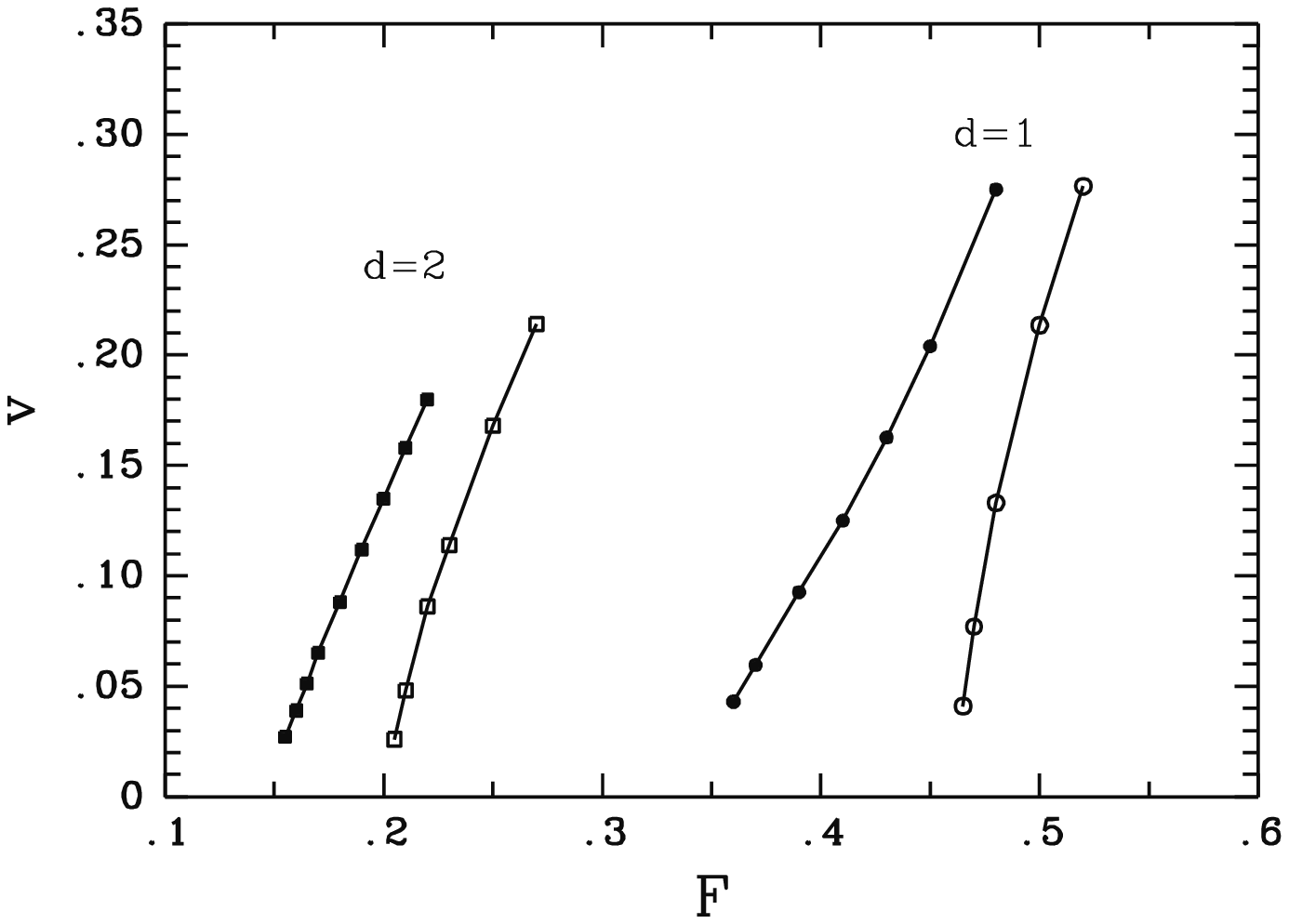}
{Average interface velocity $v$ versus the driving force $F$,
for $d=1$, $s=0$ (open circles), $d=1$, $s={1/ 2}$ (solid circles),
$d=2$, $s=0$ (open squares), and $d=2$, $s={1/ 2}$ (solid squares).}

We also measured height-height correlation functions
at the depinning transition. For a tilted surface in $d=2$,
the height fluctuations and corresponding dynamic behaviors are different
parallel and transverse to the tilt. 
The next figure shows a scaling plot of 
(a) $C_\parallel(r_\parallel,t)\equiv
\langle [h(x_\parallel+r_\parallel,x_t,t)-h(x_\parallel,x_t,t)]^2\rangle$
and (b)
$C_t(r_t,t)\equiv\langle [h(x_\parallel,x_t+r_t,t)-h(x_\parallel,x_t,t)]^2
\rangle$
against the scaled distances at the depinning threshold of
an $s={1/2}$ interface.
Each curve shows data at a given $t=32$, 64, $\cdots$, 1024, averaged
over 50 realizations of the disorder.
The data collapse is in agreement with the mapping to the KPZ equation
in one less dimension. 

\epsfysize=5cm\figure\TKDb{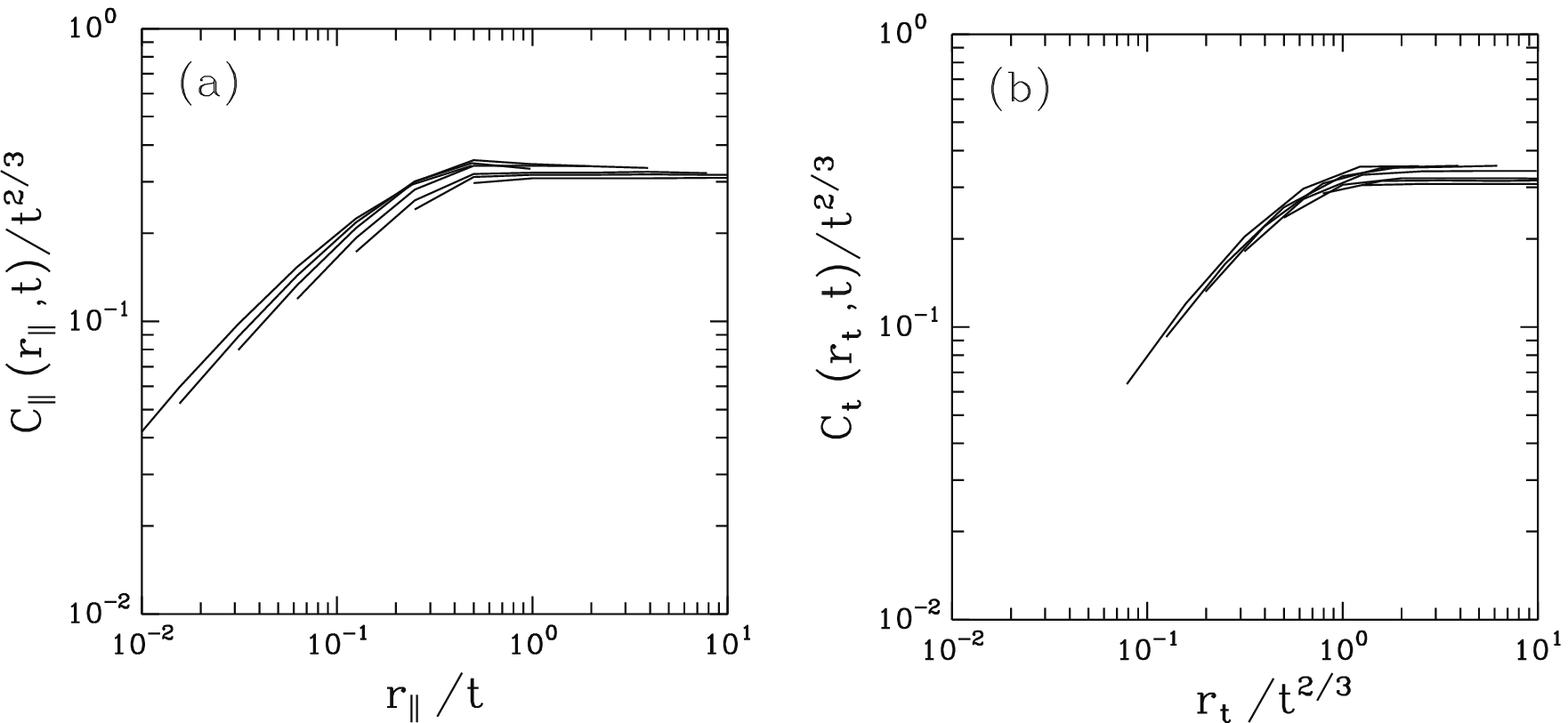}
{Height-height correlation functions (a) along and (b) transverse 
to the tilt for an $840^2$ system at different times
$32\leq t\leq 1024$. The interface at $t=0$ is flat;
$d=2$, $s={1/2}$, and $F=0.144$.}

In summary, critical behavior at the depinning of an interface depends
on the symmetries of the underlying medium. Different universality
classes can be distinguished from the dependence of the threshold
force (or velocity) on the slope, which is reminiscent of similar
dependence in a model of  resistor-diode percolation. In addition to isotropic
depinning,  we have so far identified two classes of anisotropic depinning:
along a (hard) axis of inversion symmetry in the plane, 
and tilted away from it.
We have no analytical results in the former case, but suggest a
number of scaling relations that are validated by simulations.
In the latter (more generic) case we have obtained {\it exact}
information from a mapping to moving interfaces, and confirmed
them by simulations in $d=1$ and $d=2$. As it is quite common
to encounter (intrinsic or artificially fabricated) anisotropy
for flux lines in superconductors, domain
walls in magnets, and interfaces in porous media, we expect 
our results to have important experimental ramifications.

Another form of anisotropy is also possible for interfaces in 
2+1 dimensions. If the directions $x$ and $y$ on the surface
are not related by symmetry, the non-linear term in the KPZ 
equation can be generalized, leading to the depinning equation
\eqn\edpAKPZ{\partial_t h=K_x\partial_x^2 r + K_y\partial_x^2 r +
{\lambda_x\over2} (\partial_x r)^2 + {\lambda_y\over2} (\partial_y r)^2
+ F + f(x,y,r).}
In fact the difference between $K_x$ and $K_y$ is not important
as long as both are positive. It was first pointed out by
Dietrich Wolf\ref
\rDW{D. Wolf, Phys. Rev. Lett. {\bf 67}, 1783 (1991).}\
that different signs of $\lambda_x$ and $\lambda_y$ lead to
a different universality class for the case of annealed noise. 
More recently it was demonstrated by Jeong et al\ref
\rJKK{H. Jeong, B. Kahng, and D. Kim, Phys. Rev. Lett. {\bf 77},
5094 (1996).}\
that, with quenched noise, eq.\edpAKPZ describes a new
universality class of depinning transitions with
$\beta\approx0.80(1)$, and anisotropic roughness exponents
in the $x$ and $y$ directions.

\medskip
\vskip .15truein 
\noindent{\bf 2 Fluctuating Lines}
\medskip
\noindent{\it 2.1 Flux Line Depinning}
\smallskip\noindent

The pinning of flux lines (FLs) in Type-II superconductors is of
fundamental importance to many technological applications that
require large critical currents\ref\rreview{See, for example, 
G. Blatter, M. V. Feigel'man, V. B. Geshkenbein, A. I. Larkin, and V.  
M. Vinokur, 
Rev.\ Mod.\ Phys.\ {\bf 66}, 1125 (1994); and references therein.}.
Upon application of an external current density $J$, the motion of
FLs due to the Lorentz force causes undesirable dissipation of
supercurrents. Major increases in the critical current
density $J_c$ of a sample are achieved when the FLs are pinned
to impurities. There are many recent studies,  both
experimental\ref
\rVGexp{R.~H.~Koch {\it et al.}, Phys. Rev. Lett. {\bf 63},
1511 (1989); P.~L.~Gammel, L.~F.~Schneemener, and D.J.~Bishop, 
Phys. Rev. Lett. {\bf 66}, 953 (1991).}\rsc\ref
\rcolexp{L.~Civale {\it et al.}, Phys. Rev. Lett. {\bf 67},  
648 (1991);
M.~Leghissa {\it et al.}, Phys. Rev. B {\bf 48}, 1341 (1993).}\
and theoretical\ref
\rVGtheo{D.~S.~Fisher, M.~P.~A.~Fisher, and D.~A.~Huse,
Phys. Rev. B {\bf 43}, 130 (1991).}\rsc\ref
\rcoltheo{D.~R.~Nelson and V.~M.~Vinokur, Phys. Rev. Lett.  
{\bf 68}, 2398 (1992).}, 
on collective pinning of FL's  to
point or columnar defects. Another consequence of impurities is
the strongly nonlinear behavior of the current slightly above the
depinning threshold, as the FLs start to move across the sample.
Recent numerical simulations have concentrated on the low temperature
behavior of a single FL near depinning\ref\rEnomoto{Y.~Enomoto, 
Phys. Lett. A {\bf 161}, 185 (1991); Y.~Enomoto, K.~Katsumi, R.~Kato,
and S.~Maekawa, Physica C {\bf 192}, 166 (1992).}\rsc\rDong\rsc\rFeng,
mostly ignoring fluctuations transverse
to the plane defined by the magnetic field and the Lorentz force.
Common signatures of the depinning transition from $J<J_c$ to
$J>J_c$ include a broad band ($f^{-a}$ type) voltage noise spectrum,
and self-similar fluctuations of the FL profile.

\epsfysize=7cm\figure\fFL{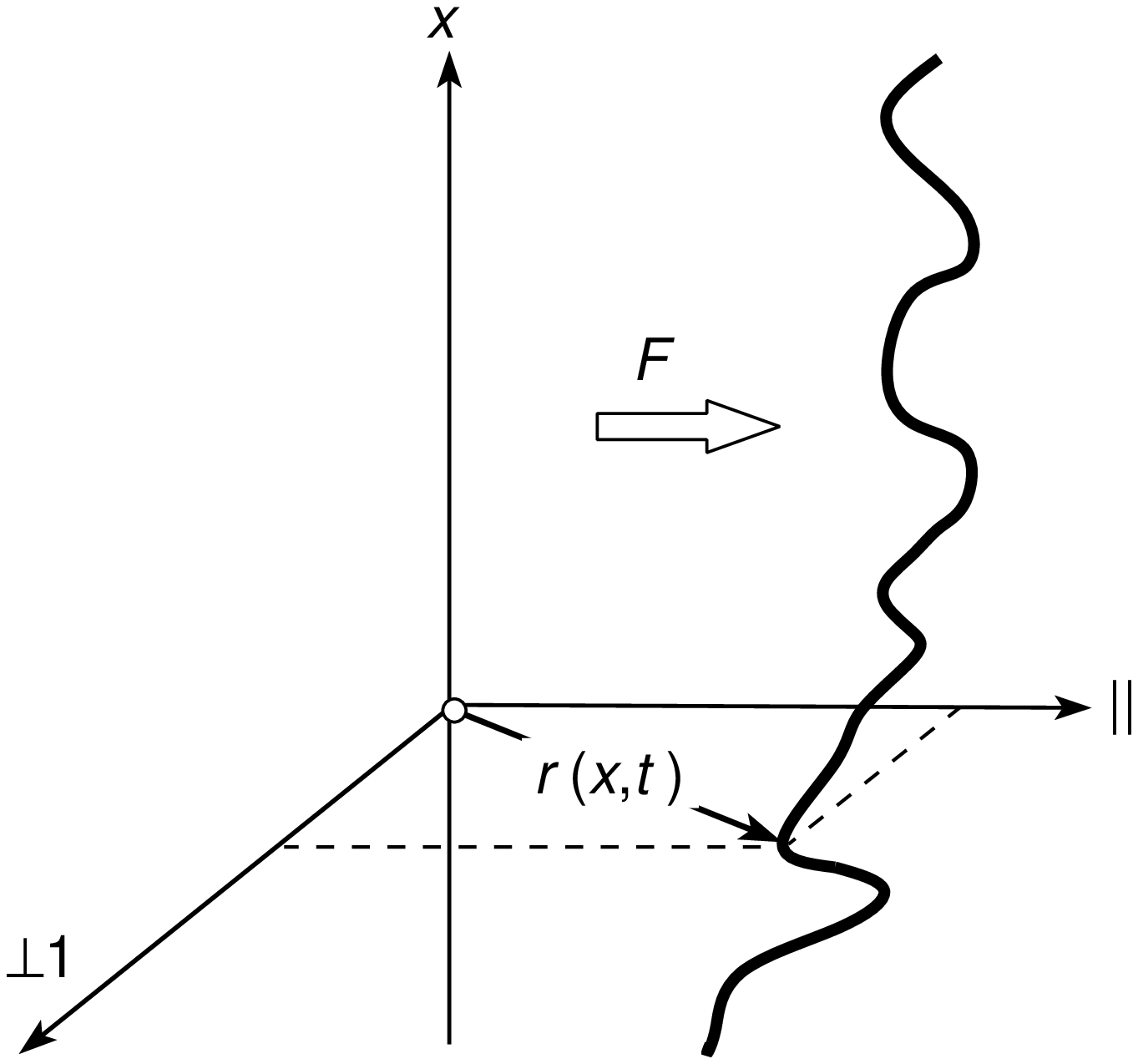}{Geometry of the line in three dimensions.}
The FL provides yet another example of a depinning transition.
We now extend the methods of the previous section to the full  
three-dimensional dynamics of a single FL at low temperatures.
The shape of the FL at a given time $t$ is described by ${\bf r}(x,t)$,
where $x$ is along the magnetic field $B$, and the unit vector
${\bf e}_\parallel$ is along the Lorentz force ${\bf F}$.
Point impurities are modeled by a random
potential $V(x,{\bf r})$, with zero mean and short-range
correlations. In the presence of impurities and a bulk Lorentz
force ${\bf F}$, the energy of a FL with small fluctuations is,
\eqn\eHamiltonian{
{\cal H} = \int dx\left\{\frac{1}{2}(\partial_x {\bf r})^2
+V\left(x,{\bf r}(x,t)\right)-{\bf r}(x,t)\cdot{\bf F}\right\}.}
The simplest possible Langevin equation for the FL, consistent with
{\it local, dissipative dynamics}, is
\eqn\emotion{
\mu^{-1}\frac{\partial {\bf r}}{\partial t} = -\frac{\delta {\cal  
H}}{\delta {\bf r}}=
\partial_x^2 {\bf r}+{\bf f} \left(x,{\bf r}(x,t)\right)+{\bf F},}
where $\mu$ is the mobility of the FL, and ${\bf f}=-\nabla_{\bf r} V$.
The potential $V(x,{\bf r})$ need not be
isotropic. For example, in a single crystal of ceramic superconductors
with the field along the oxide planes, it will be easier to move
the FL along the planes. This leads to a pinning threshold that
depends on the orientation of the force. Anisotropy also modifies the  
line tension, and the elastic term in Eq.\emotion\ is in general  
multiplied by a non-diagonal matrix $K_{\alpha\beta}$.
The random force ${\bf f}(x,{\bf r})$, can be taken to have zero
mean with correlations
\eqn\estatistics{
\langle f_\alpha(x,{\bf r})f_\gamma(x',{\bf r}')\rangle=
\delta(x-x')\Delta_{\alpha\gamma}({\bf r}-{\bf r}').}
We shall focus mostly on the isotropic case,
with $\Delta_{\alpha\gamma}({\bf r}-{\bf r}')=
\delta_{\alpha\gamma}\Delta(|{\bf r}-{\bf r}'|)$,
where $\Delta$ is a function that decays rapidly for large
values of its argument.

While the flux line is pinned by impurities when $F<F_c$,
for $F$ slightly above threshold, we expect the average velocity
$v=|{\bf v}|$ to scale as in Eq.\evelocity. Superposed on the  
steady advance of the FL are rapid  ``jumps" as portions of the line depin
from strong pinning centers. The cut off  length $\xi$ on avalanche sizes
diverges on  approaching the threshold as $\xi\sim (F-F_c)^{-\nu}$. 
At length scales up to $\xi$, the correlated fluctuations satisfy the 
dynamic scaling  forms,
\eqn\escaling{\eqalign{
\langle[r_\parallel(x,t)-r_\parallel(0,0)]^2\rangle
=& |x|^{2\zeta_\parallel}g_\parallel(t/|x|^{z_\parallel}),\cr
\langle[r_\perp(x,t)-r_\perp(0,0)]^2\rangle
=& |x|^{2\zeta_\perp}g_\perp(t/|x|^{z_\perp}),}}
where $\zeta_\alpha$ and $z_\alpha$ are the roughness and
dynamic exponents, respectively. The scaling functions
$g_\alpha$ go to a constant as their arguments approach 0.
Beyond the length scale $\xi$, different regions of the FL
depin more or less independently and  the system crosses over
to a moving state, described by different exponents, which will
be considered in the next section.

The major difference of this model from the previously studied interface
is that the position of the flux line, ${\bf r}(x,t)$, is now a
2-dimensional vector instead of a scalar; fluctuating along both
${\bf e}_\parallel$ and ${\bf e}_\perp$ directions.
One consequence is that a ``no passing" rule\ref
\rMiddleton{A.~A.~Middleton and D.~S.~Fisher, Phys. Rev. Lett.
{\bf 66}, 92 (1991); Phys. Rev. B {\bf 47}, 3530 (1993).},  
applicable to CDWs and interfaces, does not apply to FLs. 
It is possible to have coexistence of moving and stationary FLs 
in particular realizations of the random potential.
How do these transverse fluctuations scale near the depinning 
transition, and do they in turn influence the critical dynamics of 
longitudinal fluctuations near threshold?
The answer to the second question can be obtained by the following  
qualitative argument:
Consider Eq.\emotion\ for a particular realization of randomness
${\bf f}(x,{\bf r})$. Assuming that portions of the FL always
move in the forward direction, there is a unique
point $r_\perp(x,r_\parallel)$ that is visited by the line
for given coordinates $(x,r_\parallel)$. We construct a new force
field $f'$ on a two dimensional space $(x,r_\parallel)$
through $f'(x,r_\parallel)\equiv f_\parallel\left(x,r_\parallel,
r_\perp(x,r_\parallel)\right)$. It is then clear that the dynamics
of the longitudinal component $r_\parallel(x,t)$ in a given force field
${\bf f}(x,{\bf r})$ is identical to the dynamics
of $r_\parallel(x,t)$ in a force field $f'(x,r_\parallel)$, with
$r_\perp$ set to zero. It is quite plausible that, after
averaging over all ${\bf f}$, the correlations in $f'$ will
also be short-ranged, albeit different from those of
${\bf f}$. Thus, the scaling of longitudinal fluctuations of the
depinning FL will not change upon taking into account transverse
fluctuations. However, the question of how these transverse
fluctuations scale still remains.

Certain statistical symmetries of the system restrict the form of
response and correlation functions. For example, Eq.\emotion\
has statistical space- and time-translational invariance, which  
enables us to work in Fourier space, i.e. $(x,t)\to(q,\omega)$.
For an {\it isotropic} medium, ${\bf F}$ and ${\bf v}$ are parallel to
each other, i.e., ${\bf v}({\bf F})=v(F){\bf \hat F}$, where ${\bf \hat F}$  
is  the unit vector along {\bf F}. Furthermore, all
expectation values involving odd powers of a transverse
component are identically zero due to the statistical invariance  
under the transformation $r_\perp\to -r_\perp$. Thus, linear
response and two-point correlation functions  are {\it diagonal}.
The introduced critical exponents are then related through scaling  
identities. These can be derived from the linear response to an infinitesimal 
external force field $\beps(q,\omega)$,
\eqn\echi{
\chi_{\alpha\beta}(q,\omega)=\left<\frac{\partial r_\alpha(q,\omega)}
{\partial\varepsilon_\beta(q,\omega)}\right>\equiv  
\delta_{\alpha\beta}\chi_\alpha(q,\omega),}
in the $(q,\omega)\to(0,0)$ limit. Eq.\emotion\
is statistically invariant under the transformation ${\bf F}\to{\bf F}+
\varepsilon(q), \,  {\bf r}(q,\omega)\to{\bf r}(q,\omega)
+q^{-2}\varepsilon(q)$. Thus, the static linear response has the form
$\chi_\parallel(q,\omega=0)=\chi_\perp(q,\omega=0)=q^{-2}$.
Since $\varepsilon_\parallel$ scales like the applied force,
the form of the linear response at the correlation length $\xi$
gives the exponent identity
\eqn\enuexp{
\zeta_\parallel+1/\nu=2.}

Considering the transverse linear response seems to imply $\zeta_\perp=\zeta_\parallel$. 
However, the static part of the transverse linear response is irrelevant at the critical 
RG fixed point, since $z_\perp>z_\parallel$, as shown below.
When a slowly varying uniform external force 
$\varepsilon(t)$ is applied, the FL responds as if the instantaneous external force
${\bf F}+\varepsilon$ is a constant, acquiring an average velocity,
$$\langle\partial_t r_\alpha\rangle=v_\alpha({\bf F}
+\varepsilon)\approx
v_\alpha({\bf F})+\frac{\partial v_\alpha}{\partial F_\gamma}
\varepsilon_\gamma.$$
Substituting $\partial v_\parallel/\partial F_\parallel=dv/dF$ and
$\partial v_\perp/\partial F_\perp=v/F$, and Fourier transforming, gives
\eqn\echipara{\eqalign{
\chi_\parallel(q=0,\omega) =& \frac{1}{-i\omega(dv/dF)^{-1}
+O(\omega^2)}, \cr
\chi_\perp(q=0,\omega) =& \frac{1}{-i\omega(v/F)^{-1}+O(\omega^2)}.}}
Combining these with the static response, we see that the  characteristic
relaxation times of fluctuations with wavelength $\xi$ are
$$\eqalign{
\tau_\parallel(q=\xi^{-1}) \sim& \left(q^2\frac{dv}{dF}\right)^{-1}
\sim\xi^{2+(\beta-1)/\nu}\sim\xi^{z_\parallel}, \cr
\tau_\perp(q=\xi^{-1}) \sim& \left(q^2\frac{v}{F}\right)^{-1}
\sim\xi^{2+\beta/\nu}\sim\xi^{z_\perp},}$$
which, using Eq.\enuexp, yield the scaling relations
\eqn\ezperpexp{\eqalign{
\beta=(z_\parallel-\zeta_\parallel)\nu, \cr
z_\perp=z_\parallel+1/\nu.}}
We already see that the dynamic relaxation of transverse
fluctuations is much slower than longitudinal ones.
All critical exponents can be calculated from $\zeta_\parallel$,
$\zeta_\perp$, and $z_\parallel$, by using Eqs\enuexp,  and \ezperpexp.

Equation \emotion\ can be analyzed using the formalism of
Martin, Siggia, and Rose (MSR)\rMSR. Ignoring transverse
fluctuations, and generalizing to  $d$ dimensional internal coordinates
${\bf x}\in\Re^d$, leads to an interface
depinning model which was studied by Nattermann, Stepanow, Tang, and
Leschhorn (NSTL)\rNSTL, and by Narayan and Fisher
(NF)\rNF. The RG treatment indicates that impurity
disorder becomes relevant for dimensions $d\leq4$, and the
critical exponents in $d=4-\epsilon$ dimensions are given to one-loop
order as $\zeta=\epsilon/3$, $z=2-2\epsilon/9$.
NSTL obtained this result by directly averaging the MSR
generating functional $Z$, and calculating the renormalization
of the force-force correlation function $\Delta(r)$, perturbatively
around the freely moving interface $[\Delta(r)=0]$.
NF, on the other hand, used a perturbative expansion of $Z$, around a
saddle point corresponding to a mean-field approximation to
Eq.\emotion\ref
\rSZ{H.~Sompolinsky and A.~Zippelius, Phys. Rev. B {\bf 25},
6860 (1982); A.~Zippelius,  Phys. Rev. B {\bf 29}, 2717 (1984).}, 
which involved {\it temporal}
force-force correlations $C(vt)$. They argue that a
conventional low-frequency analysis is not sufficient to determine
critical exponents. They  also suggest that the roughness
exponent is equal to $\epsilon/3$ to all orders in perturbation
theory.

Following the approach of NF, we employ a perturbative expansion
of the disorder-averaged MSR partition function around a mean-field
solution for cusped impurity potentials\rNF. All terms in the
expansion involving longitudinal fluctuations are identical to the
interface case, thus we obtain the same critical exponents for
longitudinal fluctuations, i.e., $\zeta_\parallel=\epsilon/3$,
$z_\parallel=2-2\epsilon/9+O(\epsilon^2)$. Furthermore, {\it for isotropic
potentials}, the renormalization of transverse temporal force-force
correlations $C_\perp(vt)$ yields a transverse roughness exponent
$\zeta_\perp=5\zeta_\parallel/2-2$, to all orders in perturbation
theory. For the FL $(\epsilon=3)$, the critical exponents are then given by
\eqn\eIII{\eqalign{
\zeta_\parallel=1,\quad z_\parallel\approx4/3,\quad &\nu=1, \cr
\beta\approx1/3,\quad \zeta_\perp=1/2,\quad &z_\perp\approx7/3.}}

\epsfysize=5cm\figure\fVE{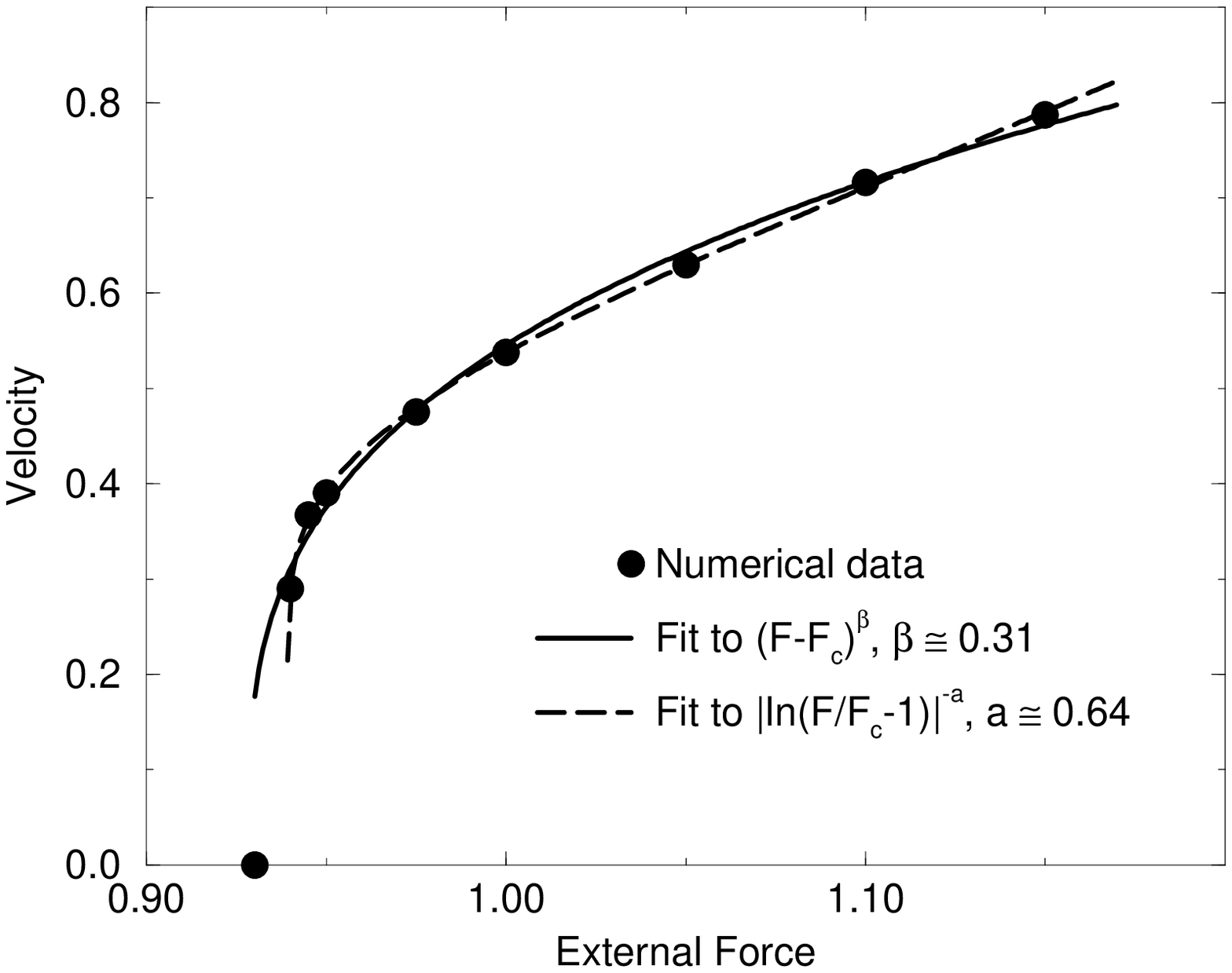}{A plot of average velocity versus 
external force for a system of 2048 points. Statistical errors are smaller 
than symbol sizes. Both fits have three adjustable parameters: The
threshold force, the exponent, and an overall multiplicative constant.}

\epsfysize=5cm\figure\fRE{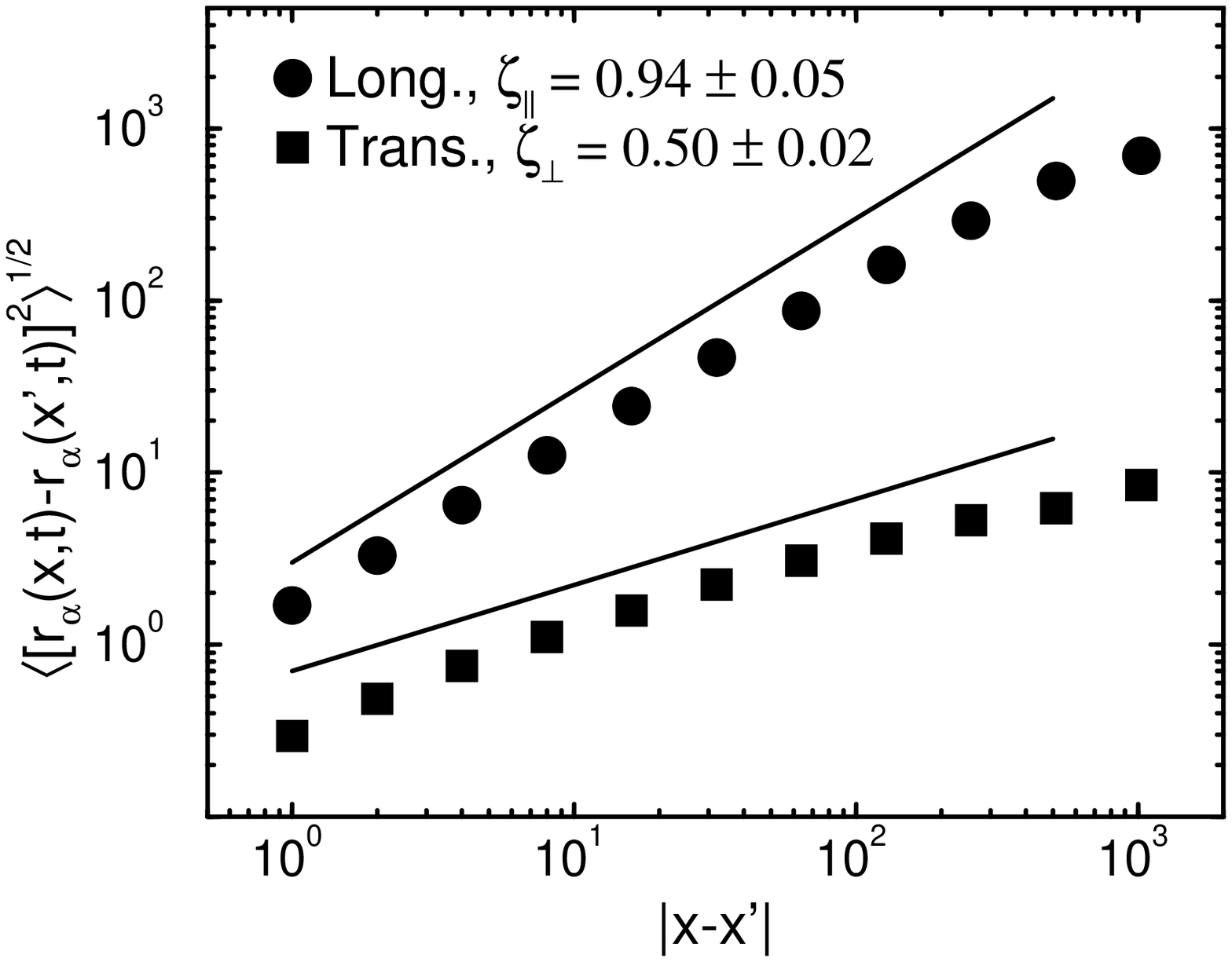}{A plot of equal time correlation 
functions versus separation, for the system shown in Fig.\fVE, 
at $F=0.95$. 
The observed roughness exponents very closely follow the theoretical 
predictions of $\zeta_\parallel=1,\;\zeta_\perp=0.5$, which are shown 
as solid lines for comparison.}

To test the scaling forms and  exponents predicted by Eqs.\evelocity\
and \escaling, we numerically integrated Eq.\emotion,
discretized in coordinates $x$ and $t$.
Free boundary conditions were used for system sizes of up to
2048, with a grid spacing $\Delta x=1$ and a time step $\Delta  t=0.02$.
Time averages were evaluated {\it after}
the system reached steady state. Periodic boundary conditions gave
similar results, but with larger finite size effects. Smaller grid  sizes
did not change the results considerably.
The behavior of $v(F)$  seems to fit the scaling form of
Eq.\evelocity\ with an exponent $\beta\approx0.3$, but is also
consistent with a logarithmic dependence on the reduced force, i.e.,
$\beta=0$. The same behavior was observed by Dong {\it et al.}
in a recent simulation of the $1+1$ dimensional geometry\rDong.
Since $z_\parallel$, and consequently $\beta$, is known only to
first order in $\epsilon$, higher order corrections are expected.
By looking at equal time correlation functions, we  
find that transverse fluctuations are strongly suppressed, and that the roughness
exponents are equal to our theoretical estimates within statistical
accuracy. The excellent agreement for
$\epsilon=3$ suggests that the theoretical estimates are indeed  exact.

The potential pinning the FL in a single superconducting crystal is
likely to be highly {\it anisotropic}. For example, consider a  
magnetic field parallel to the copper oxide planes of a ceramic  
superconductor. The threshold force then depends on its orientation,
with depinning easiest along the copper oxide planes.
In general, the average velocity may depend on the orientations
of the external force and the FL. The most general gradient
expansion for the equation of motion is then,
\eqn\eAnis{
\frac{\del r_\alpha}{\del t} = \mu_{\alpha\beta}F_\beta +
\kappa_{\alpha\beta}\del_x r_\beta +K_{\alpha\beta}\del_x^2 r_\beta
+\frac{1}{2}\Lambda_{\alpha,\beta\gamma}\del_xr_\beta\del_xr_\gamma
+f_\alpha\left(x,\br(x,t)\right)+\cdots,}
with
\eqn\eAnisnoise{
\langle f_\alpha(x,\br)f_\beta(x',\br')\rangle 
= \delta(x-x')C_{\alpha\beta}(\br-\br').}
Depending on the presence or absence of various terms allowed by
the symmetries of the system, the above set of equations encompasses
many distinct universality classes. For example, consider the situation
where \bv\ depends on \bF, but not on the orientation of the line. 
Eqs.\echipara\ have to be modified, since \bv\ and \bF\ are no
longer parallel (except along the axes with
$\br\to -\br$ symmetry), and the linear response function is not
diagonal. The RG analysis is more cumbersome:
For depinning along a non-symmetric direction,
the longitudinal exponents are not modified (in agreement
with the argument presented earlier), while the transverse
fluctuations are further suppressed to $\zeta_\perp=
2\zeta_\parallel-2$ (equal to zero for  
$\zeta_\parallel=1$)\ref\rfootnote{In this case, 
the longitudinal direction is chosen to be
along the average velocity \bv, not the Lorentz force \bF.}.
Relaxation of transverse modes are still characterized by
$z_\perp=z_\parallel+1/\nu$, and the exponent identity \enuexp\
also holds. Surprisingly, the exponents for depinning along
axes of reflection symmetry are the same as the isotropic case.

If the velocity also depends on the tilt, there will be 
additional relevant terms in the MSR partition function, which
invalidate the arguments leading to Eq.\enuexp.
The analogy to FLs in a planes suggests that the longitudinal exponents 
for $d=1$ are controlled by DP clusters\rTL$^,$\rBul, with 
$\zeta_\parallel\approx 0.63$. Since no
perturbative fixed point is present in this case, it is not
clear how to explore the behavior of transverse fluctuations
systematically.

\medskip
\noindent{\it 2.2 Dynamic Fluctuations of an Unpinned Flux Line}
\smallskip\noindent

So far,w investigated the dynamics of a Flux Line near the depinning 
transition. Now, we would like to consider its behavior in a different
regime, when the external driving force is large, and the impurities
appear as weak barriers that deflect portions of the line
without impeding its overall drift. In such non--equilibrium systems, 
one can regard the evolution equations as more fundamental,
and proceed by constructing the most general equations consistent with the
symmetries and conservation laws of the situation under study\ref
\rMK{M. Kardar, in {\it Disorder and Fracture}, edited by J.C. Charmet, 
S. Roux, and E. Guyon, Plenum, New York (1990); T. Hwa and M. Kardar, Phys.
Rev. A {\bf 45}, 7002 (1992).}.
Even in a system with isotropic randomness, which we will discuss here,
the average drift velocity, $v$, breaks the symmetry between
forward and backward motions, and allows introduction of 
nonlinearities in the equations of motion\ref
\rPRL{M. Plischke, Z. R\'acz, and D. Liu, Phys. Rev. B {\bf 35}, 3485 (1987).}
$^,$\rMK. 

Let us first concentrate on an interface in two dimensions. (Fig.\CL.)
By contracting up to two spatial derivatives of $r$, and keeping terms
that are relevant, one obtains 
the Kardar-Parisi-Zhang\rKPZ\ (KPZ) equation,
\eqn\eKPZ{
\del_tr(x,t)=\mu F+K\del_x^2r(x,t)+\frac{\lambda}{2}
\left[\del_xr(x,t)\right]^2+f(x,t),}
with random force correlations
\eqn\enoise{\langle f(x,t)f(x',t')\rangle=
2T\delta(x-x')\delta(t-t').}
For a moving line, the term proportional to the external force can
be absorbed without loss of generality by considering a suitable 
Galilean transformation, $r\to r-at$, to a moving frame.
A large number of stochastic nonequilibrium growth models,
like the Eden Model and various ballistic deposition models 
are known to be well described, at large length scales and times, 
by this equation, which is intimately related to several other
problems. For example,
the transformation $v(x,t)=-\lambda \del_x r(x,t)$ maps Eq.\eKPZ\
to the randomly stirred {\it Burgers' equation} for fluid flow\ref
\rJMB{J.M. Burgers, {\it The Nonlinear Diffusion Equation} (Riedel,
Boston, 1974).}\rsc\ref\rFNS{D.~Forster, D.~R.~Nelson, and 
M.~J.~Stephen, Phys. Rev. A {\bf 16}, 732 (1977).},
\eqn\eBurg{\del_tv+v\del_xv=K\del_x^2v-\lambda\del_xf(x,t).}

The correlations of the line profile still satisfy 
the dynamic scaling form in Eq.\eHcorr, nevertheless with different
scaling exponents $\zeta, z$ and scaling function $g$. This self-affine
scaling is not critical, i.e., not obtained by fine tuning an external
parameter like the force, and is quite different in nature than the 
critical scaling of the line near the depinning transition, which 
ceases beyond the correlation length scale $\xi$.

Two important nonperturbative properties of Eq.\eKPZ\ help us determine
these exponents exactly in 1+1 dimensions:

\noindent{\bf 1.} {\it Galilean Invariance (GI):} Eq.\eKPZ\ is statistically invariant 
under the infinitesimal reparametrization
\eqn\eGI{r'=r+\epsilon x~,~x'=x+\lambda\epsilon t~,~t'=t,}
provided that the random force $f$ does not have temporal correlations\ref
\rMHKZ{E.~Medina, T.~Hwa, M.~Kardar, and Y.~Zhang, Phys. Rev. A
{\bf 39}, 3053 (1989).}.
Since the parameter $\lambda$ appears both in the transformation and 
Eq.\eKPZ, it is not renormalized under any RG procedure that preserves
this invariance.
This implies the exponent identity\rFNS$^,$\rMHKZ
\eqn\eGIid{\zeta+z=2.}

\noindent{\bf 2.} {\it Fluctuation--Dissipation (FD) Theorem:} Eqs.\eKPZ\ and \enoise\
lead to a Fokker--Planck equation for the evolution of
the joint probability ${\cal P}\left[r(x)\right]$,
\eqn\eFP{
\partial_t{\cal P}=\int dx\,\left( {\delta {\cal P}\over
\delta r(x)}\,\partial_t r + T{\delta^2{\cal P}
\over [\delta r(x)]^2}\right).
}
It is easy to check that ${\cal P}$ has a stationary solution
\eqn\eSS{{\cal P}=\exp\left(-{K\over 2T}\int dx\,(\del_x r)^2\right).}
If ${\cal P}$ converges to this solution, 
the long--time behavior of the correlation
functions in Eq.\eHcorr\ can be directly read off Eq.\eSS, 
giving $\zeta=1/2$.

Combining these two results, the roughness and dynamic exponents are
exactly determined for the line in two dimensions as
\eqn\eKPZexp{\zeta=1/2~,\qquad z=3/2.}
Many direct numerical simulations and discrete growth models have 
verified these exponents to a very good accuracy. Exact exponents 
for the {\it isotropic} KPZ equation are
not known in higher dimensions, since the FD property
is only valid in two dimensions. These results have been summarized in
a number of recent reviews\ref\rFV{{\it Dynamics of Fractal Surfaces}, 
edited by F. Family and T. Vicsek (World Scientific, 
Singapore, 1991).}$^,$\ref\rKS{J.~Krug and H.~Spohn, in {\it Solids 
Far From Equilibrium: Growth, Morphology and Defects}, edited by 
 C.~Godreche (Cambridge University Press, Cambridge, 1991).}\rsc\ref
\rHHZ{T. Halpin--Healy and Y.-C. Zhang, Phys. Rep. {\bf 254}, 215 (1995).}\rsc\ref
\rBS{A.-L. Barabasi and H. E. Stanley, {\it Fractal concepts in surface growth},
(CUP, Cambridge, 1995).}. 

As an aside we remark that some exact information is available for the
{\it anisotropic} KPZ equation in 2+1 dimensions. Using a perturbative
RG approach, Wolf showed\rDW\ that in the equation
\eqn\eAKPZ{\partial_t r=K\nabla^2 r +
{\lambda_x\over2} (\partial_x r)^2 + {\lambda_y\over2} (\partial_y r)^2
+ f(x,y,t),}
the nonlinearities $\{\lambda_x,\lambda_y\}$ renormalize to zero if
they initially have opposite signs. This suggests logarithmic fluctuations
for the resulting interface, as in the case of the linear Langevin equation.
In fact, it is straightforward to demonstrate that eq.\eAKPZ\ also
satisfies a Fluctuation Dissipation condition if $\lambda_x=-\lambda_y$.
When this condition is satisfied, the associated Fokker--Planck
equation has a steady state solution
\eqn\eSSAKPZ{{\cal P}=\exp\left(-{K\over 2T}
\int dxdy\,(\nabla r)^2\right).}
This is a non--perturbative result which again indicates the logarithmic
fluctuations resulting from eq.\eAKPZ\. In this context, it is interesting
to note that the steady state distribution for an exactly solvable discrete 
model of  surface growth belonging to the above universality class
has also been obtained\ref
\rPS{M. Prahofer and H. Spohn, J. Stat. Phys., in press (1997).}

Let us now turn to the case of a line in three dimensions (Fig.\fFL ). 
Fluctuations of the line can be indicated by a
a two dimensional vector $\br$. Even in an isotropic medium, 
the drift velocity $\bv$ breaks the isotropy in \br\ by selecting 
a direction. A gradient expansion up to second order
for the equation of motion  
gives\ref\rEKlines{D.~Ertas and M.~Kardar, 
Phys. Rev. Lett. {\bf 69}, 929 (1992).}
\eqn\eGrow{\eqalign{
\del_t r_\alpha=&\left[ K_1 \delta_{\alpha\beta}+
K_2 v_\alpha v_\beta\right] \del_x^2 r_\beta \cr
\noalign{\medskip}
&\quad+\left[ \lambda_1 (\delta_{\alpha\beta}v_\gamma 
+\delta_{\alpha\gamma}v_\beta)
+\lambda_2 v_\alpha \delta_{\beta\gamma} + \lambda_3 v_\alpha 
v_\beta v_\gamma
\right] {\del_x r_\beta \del_x r_\gamma \over 2}+ f_\alpha},}
with random force correlations
\eqn\eTT{\langle f_\alpha(x,t)f_\beta(x',t')\rangle=
2[T_1\delta_{\alpha\beta}+T_2v_\alpha v_\beta]\delta(x-x')\delta(t-t').} 
Higher order nonlinearities can be similarly constructed but are in fact 
irrelevant.
In terms of components parallel and perpendicular to the
velocity, the equations are
\eqn\eGro{\left\{\eqalign{\del_t \hl &= \Kl\del_x^2 \hl 
+ {\lal  \over 2} (\del_x \hl)^2+ {\lalt \over 2}(\del_x \htr)^2
+\fl(x,t) \cr \noalign{\medskip} 
\del_t\htr&=\Kt\del_x^2 \htr+\lat\del_x\hl\del_x \htr+\ft(x,t)} \right.
~~,}
with
\eqn\eD{\left\{\eqalign{\langle \fl(x,t)\fl(x',t')\rangle=&
2\Tl\delta(x-x')\delta(t-t') \cr \noalign{\medskip}
\langle \ft(x,t)\ft(x',t')\rangle=&
2\Tt\delta(x-x')\delta(t-t')}\right.~~.}
The noise-averaged correlations have a dynamic scaling form like 
Eq.\escaling,
\eqn\eCor{\left\{\eqalign{
\langle[r_\parallel(x,t)-r_\parallel(x',t')]^2\rangle
&= |x-x'|^{2\zeta_\parallel}g_\parallel\left(
\frac{|t-t'|}{|x-x'|^{z_\parallel}}\right),\cr
\langle[r_\perp(x,t)-r_\perp(x',t')]^2\rangle
&= |x-x'|^{2\zeta_\perp}g_\perp\left(
\frac{|t-t'|}{|x-x'|^{z_\perp}}\right).\cr }\right.
}

\epsfysize10cm\figure\fRGflow{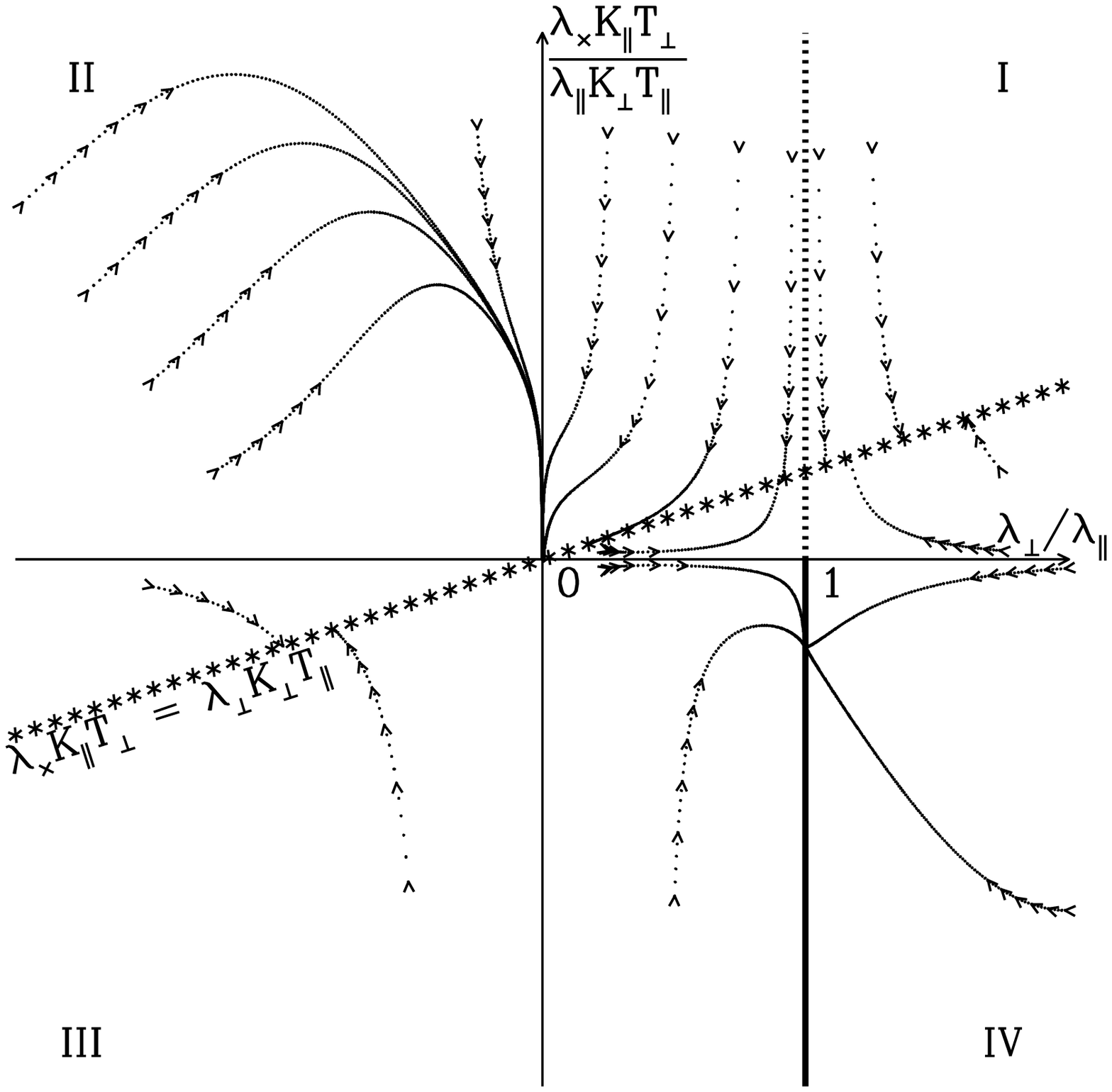}{A projection of RG flows 
in the parameter space, for $n=1$ transverse components.}

In the absence of nonlinearities $(\lal=\lalt=\lat=0)$, Eqs.\eGro\
can easily be solved to give
$\zetal=\zetat=1/2$ and $\zl=\zt=2$. Simple dimensional counting
indicates that all three nonlinear terms are relevant 
and may modify the exponents in Eq.\eCor.
Studies of related stochastic equations\ref\rTH{T.~Hwa, Phys. Rev. Lett.
{\bf 69}, 1552 (1992).}$^,$\rDW\
indicate that interesting dynamic phase diagrams may emerge from
the competition between nonlinearities.
Let us assume that $\lal$ is positive and finite (its sign can be 
changed by $\hl\to -\hl$), and focus on the dependence
of the scaling exponents on the ratios $\lat/\lal$ and $\lalt/\lal$, as
depicted in Fig.\fRGflow. (It is more convenient to set
the vertical axis to $\lalt\Kl\Tt/\lal\Kt\Tl$.) 

The properties discussed for the KPZ equation can be extended to this
higher dimensional case:

\noindent{\bf 1.} {\it Galilean Invariance (GI):} Consider the infinitesimal 
reparametrization
\eqn\eGal{\left\{\eqalign{x'=x+\lal\epsilon t ~&,~t'=t~,\cr
\hl'=\hl+\epsilon x ~&,~\htr'=\htr~.}\right.}
Eqs.\eGro\ are invariant under this transformation
provided that $\lal=\lat$. 
Thus {\it along this line} in Fig.\fRGflow\ there is GI, which 
once more implies the exponent identity
\eqn\eGInew{\zetal+\zl=2.}

\noindent{\bf 2.} {\it Fluctuation--Dissipation (FD) Condition:} 
The Fokker--Planck 
equation for the evolution of the joint probability 
${\cal P}\left[ \hl(x),\htr(x)\right]$ has a stationary solution
\eqn\eSS{{\cal P}_0\propto\exp\left(-\int dx\left[{\Kl\over 2\Tl}(\del_x\hl)^2
+{\Kt\over 2\Tt}(\del_x\htr)^2\right]\right),}
provided that $\lalt\Kl\Tt=\lat\Kt\Tl$. Thus for this special choice of 
parameters, depicted by a starred line in Fig.\fRGflow, if ${\cal P}$
converges to this solution, the long--time behavior of the correlation
functions in Eq.\eCor\ can be directly read off Eq.\eSS, 
giving $\zetal=\zetat=1/2$. 

\noindent{\bf 3.} {\it The Cole--Hopf (CH) Transformation} is an important 
method for the exact study of solutions of the one component 
nonlinear diffusion 
equation\rJMB. Here we generalize this transformation to the complex 
plane by defining, {\it for }$\lalt<0$,
\eqn\eCH{
\Psi(x,t)=\exp\left(\frac{\lal \hl(x,t)+i\sqrt{-\lal\lalt}\htr(x,t)}{2K}
\right).}
The linear diffusion equation 
$$\del_t \Psi=K\del_x^2 \Psi +\mu(x,t)\Psi,$$

\noindent then leads to Eqs.\eGro\ if $\Kl=\Kt=K$ and $\lal=\lat$. 
[Here ${\rm Re}(\mu)=\lal\fl/2K$ and ${\rm Im}(\mu)=
\sqrt{-\lal\lalt}\ft/2K$.]
This transformation enables an exact solution of the
{\it deterministic} equation, and further allows us to write the solution
to the {\it stochastic} equation in the form of a path integral
\eqn\ePath{\Psi(x,t)=\int_{(0,0)}^{(x,t)}{\cal D}x(\tau)
\exp  \left\{ -\int_0^t d\tau  \left[{{\dot x}^2 \over 2K} 
+\mu(x,\tau)\right]  \right\} .}
Eq.\ePath\ has been extensively studied in connection with quantum
tunneling in a disordered medium\ref
\rMKSW{E. Medina, M. Kardar, Y. Shapir, and X.-R. Wang, Phys. Rev. Lett. 
{\bf 62}, 941 (1989); E. Medina and M. Kardar, Phys. Rev. B {\bf 46}, 9984 (1992).},
with $\Psi$ representing the wave function. In particular, results for the
tunneling probability $|\Psi|^2$ suggest $\zl=3/2$ and
$\zetal=1/2$. The transverse fluctuations correspond to the phase
in the quantum problem which is not an observable. Hence this mapping does
not provide any information on $\zetat$ and $\zt$ which
are in fact observable for the moving line.

At the point $\lat=\lalt=0$, $\hl$ and $\htr$ decouple, 
and $\zt=2$ while $\zl=3/2$. However, in general 
$\zl=\zt=z$ unless the effective $\lat$ is zero. For example
at the intersection of the subspaces with GI and FD the 
exponents $\zl=\zt=3/2$ are obtained from the 
exponent identities. Dynamic RG recursion relations 
can be computed to one--loop order\rEKlines$^,$\ref\rEKpoly{D.~Ertas and 
M.~Kardar, Phys. Rev. E {\bf 48}, 1228 (1993).},
by standard methods of momentum-shell dynamic RG\rFNS$^,$\rMHKZ.

The renormalization of the seven parameters in Eqs.\eGro, generalized
to $n$ transverse directions, give the recursion relations
\eqn\eRR{\eqalign{
{d K_\parallel\over d\ell} &=   
 K_\parallel\left[z-2+{1\over\pi}{\lambda_\parallel^2T_\parallel\over 
4 K_\parallel^3}
+n{1\over\pi}{\lambda_\perp\lambda_\times T_\perp\over 4 K_\parallel 
K_\perp^2}\right],\cr
{d K_\perp \over d\ell} &= 
 K_\perp\left[z-2+{1\over\pi}{\lambda_\perp\big( (\lambda_\times 
T_\perp/ K_\perp)
+(\lambda_\perp T_\parallel/ K_\parallel)\big) \over 2 K_\perp 
( K_\perp+ K_\parallel)} \right. \cr
 & \hskip 1in \left. +{1\over\pi}{ K_\perp- K_\parallel \over  K_\perp+ K_\parallel}
{\lambda_\perp\big( (\lambda_\times T_\perp/ K_\perp)
-(\lambda_\perp T_\parallel/ K_\parallel)\big) \over  K_\perp 
( K_\perp+ K_\parallel)}\right],\cr
{dT_\parallel \over d\ell} &=  T_\parallel\left[z-2\zeta_\parallel-1
+{1\over\pi}{\lambda_\parallel^2T_\parallel \over 4 K_\parallel^3}
\right]+n{1\over\pi}{\lambda_\times^2T_\perp^2 \over 4 K_\perp^3},\cr
{dT_\perp \over d\ell} &= T_\perp\left[z-2\zeta_\perp-1
+{1\over\pi}{\lambda_\perp^2T_\parallel \over  K_\perp K_\parallel
( K_\perp+ K_\parallel)}\right],\cr
{d\lambda_\parallel \over d\ell}  &= \lambda_\parallel\left[
\zeta_\parallel+z-2\right],\cr
{d\lambda_\perp \over d\ell}  &= \lambda_\perp\left[
\zeta_\parallel+z-2-{1\over\pi}
{\lambda_\parallel-\lambda_\perp \over ( K_\perp+ K_\parallel)^2}
{\big( (\lambda_\times T_\perp/ K_\perp)-(\lambda_\perp  
T_\parallel/ K_\parallel)\big)}\right],\cr
{d\lambda_\times \over d\ell} &= \lambda_\times\left[
2\zeta_\perp-\zeta_\parallel+z-2
+{1\over\pi}{\lambda_\parallel K_\perp-\lambda_\perp K_\parallel 
\over  K_\perp K_\parallel( K_\perp+ K_\parallel)}
\big( (\lambda_\times T_\perp/ K_\perp)\right. \cr
 & \hskip 1in \left. -(\lambda_\perp  
T_\parallel/ K_\parallel)\big)\right]. \cr
}}

The projections of the RG flows on the two parameter subspace shown in 
Fig.\fRGflow\ are indicated by trajectories. 
They naturally satisfy the constraints
imposed by the non--perturbative results: the subspace of GI is closed under
RG, while the FD condition appears as a {\it fixed line}. The RG
flows, and the corresponding exponents, are different in each quadrant 
of Fig.\fRGflow, which implies that the scaling behavior is determined
by the relative signs of the three nonlinearities. This was confirmed by
numerical integrations\rEKlines$^,$\rEKpoly\ of  
Eqs.\eGro,  performed for different sets of parameters. A summary of the 
computed exponents are given in Table I.

The analysis of analytical and numerical results can be summarized
as follows:

$\lat\lalt>0$ : In this region, the scaling behavior is understood best. 
The RG flows terminate on the fixed line where FD conditions
apply, hence $\zetal=\zetat=1/2$. All along this line, the one
loop RG exponent is $z=3/2$. These results are consistent with the numerical
simulations. The measured exponents rapidly converge to these values, except
when $\lat$ or $\lalt$ are small.

$\lalt=0$: In this case the equation for $\hl$ is the KPZ equation \eKPZ,
thus $\zetal=1/2$ and $\zl=3/2$. The fluctuations in 
$\hl$ act as a strong (multiplicative and correlated) noise on $\htr$.
The one--loop RG yields the exponents 
$\zt=3/2,\ \zetat=0.75$\ for $\lat>0$, while 
a negative $\lat$ scales to 0 suggesting $z_\perp>z_\para$.
Simulations are consistent with the RG calculations for $\lat>0$, yielding
$\zeta_\perp=0.72$, surprisingly close to the one--loop RG value.
For $\lat<0$, simulations indicate $z_\perp\approx 2$ and 
$\zeta_\perp\approx 2/3$ along with the expected values for the 
longitudinal exponents.

$\lat=0$: The transverse fluctuations satisfy a simple diffusion equation
with $\zetat=1/2$ and $\zt=2$. Through the term $\lalt(\del_x \htr)^2/2$,
these fluctuations act as a correlated noise\rMHKZ\ for the longitudinal 
mode. A naive application of the results of this  reference\rMHKZ\  give
$\zetal=2/3$ and $\zl=4/3$. 
Quite surprisingly, simulations indicate different behavior depending on
the sign of $\lalt$. 
For $\lalt<0$, $\zl\approx 3/2$ and $\zetal\approx1/2$
whereas for $\lalt>0$, longitudinal fluctuations are much stronger,
resulting in $\zl\approx 1.18$ and $\zetal\approx 0.84$. Actually,
$\zetal$ increases steadily with system size, suggesting a breakdown 
of dynamic scaling, due to a change of sign in $\lat\lalt$. 
This dependence on the sign of $\lalt$ may reflect
the fundamental difference between behavior
in quadrants II and IV of Fig.\fRGflow.

\epsfysize=10cm
\medskip
\centerline{\epsffile{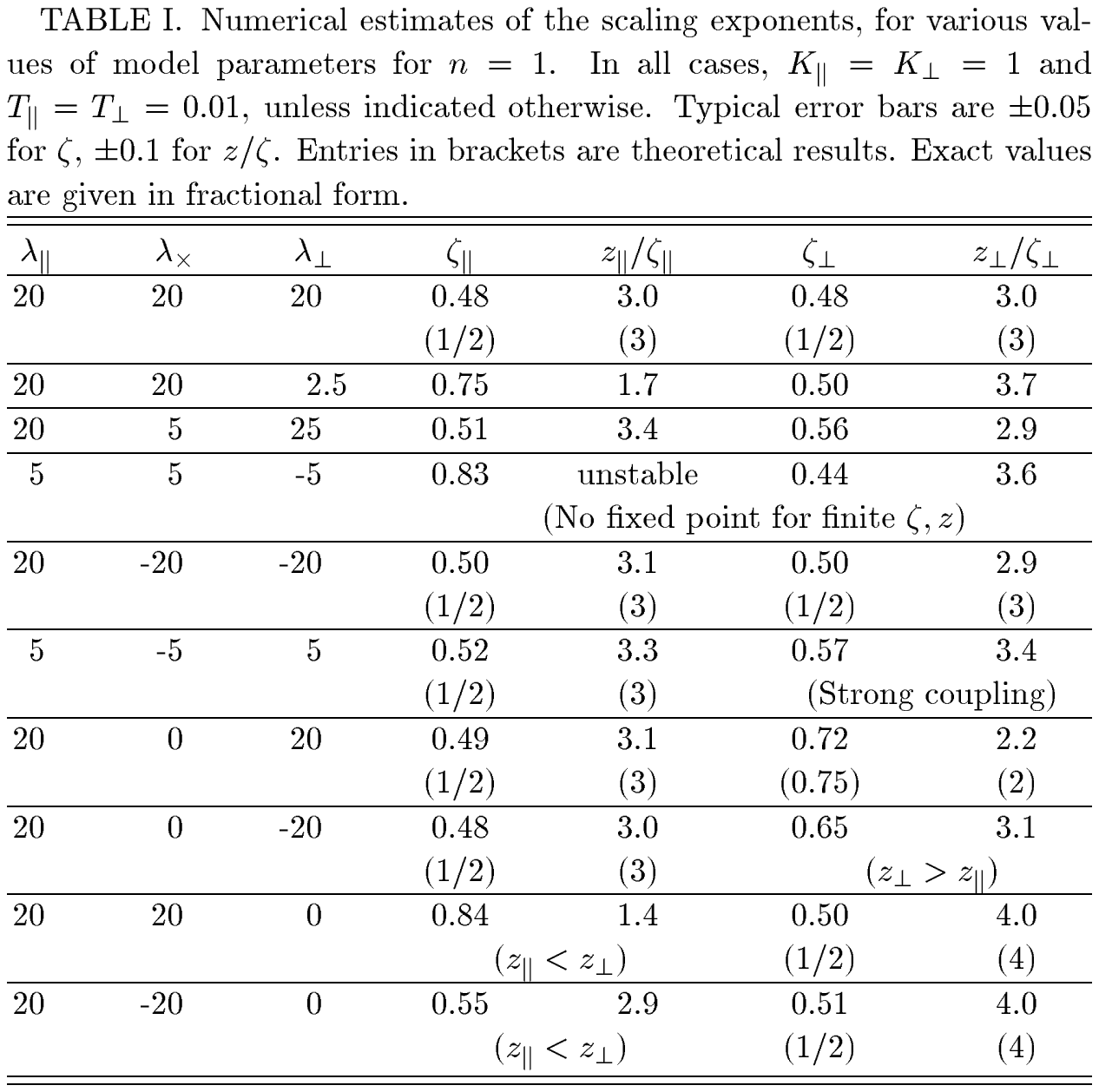}}
\medskip

$\lat<0$ {\it and} $\lalt>0$: The analysis of this region (II) is the most
difficult in that the RG flows do not converge upon a finite fixed point 
{\it and } $\lat\to 0$, which may signal the breakdown of dynamic scaling.
Simulations indicate strong longitudinal fluctuations that lead
to instabilities in the discrete integration scheme, excluding the 
possibility of measuring the exponents reliably. 

$\lat>0$ {\it and} $\lalt<0$: The {\it projected\/} RG flows in this 
quadrant (IV) converge to the point $\lat/\lal=1$ and $\lalt\Tt\Kl/
\lal\Tl\Kt=-1$. This is actually not a fixed point, as
$K_\para$ and $K_\perp$ scale to infinity. The applicability
of the CH transformation to this point implies $\zl=3/2$ and $\zetal=1/2$. 
Since $\lat$ is finite, $\zt=\zl=3/2$ is expected, but this does
not give any information on $\zetat$. Simulations indicate strong 
transverse fluctuations and suffer from difficulties similar to
those in region II.

Eqs.\eGro\ are the simplest nonlinear, local, 
and dissipative equations that govern the fluctuations of a moving
line in a random medium. They can be easily generalized to describe 
the time evolution of a manifold with arbitrary internal 
($\bx \in R^{d}$) and external 
($\br \in R^{n+1}$) dimensions, and to the motion of curves
that are not necessarily stretched in a particular direction. 
Since the derivation only involves
general symmetry arguments, the given results are widely applicable
to a number of seemingly unrelated systems. 
We will discuss one application to drifting polymers in more
detail in the next lecture, explicitly demonstrating the origin
of the nonlinear terms starting from more fundamental 
hydrodynamic equations. A simple model of crack front propagation
in three dimensions\ref\rBBLP{J.~P.~Bouchaud, E.~Bouchaud, G.~Lapasset,
and J.~Planes, Phys. Rev. Lett. {\bf 71}, 2240 (1993).}\ 
also arrives at Eqs.\eGro, implying the self-affine
structure of the crack surface after the front has passed. 

\medskip
\noindent{\it 2.3 Drifting Polymers}
\smallskip\noindent

The dynamics of polymers in fluids is of much theoretical interest
and has been extensively studied\ref\rDoi{M.~Doi and S.F.~Edwards, 
{\it Theory of Polymer Dynamics}, 
Oxford University Press (1986).}$^,$\ref\rDeG{P.G.~de~Gennes, 
{\it Scaling Concepts in Polymer Physics},
Cornell University Press (1979).}. The combination
of polymer flexibility, interactions, and hydrodynamics make a
first principles approach to the problem quite difficult. There are,
however, a number of phenomenological studies that describe various
aspects of this problem\ref\rBrd{R.B.~Bird, {\it Dynamics of Polymeric Physics},
Vols. 1-2,
Wiley, New York (1987).}.
\epsfysize=7cm\figure\fPoly{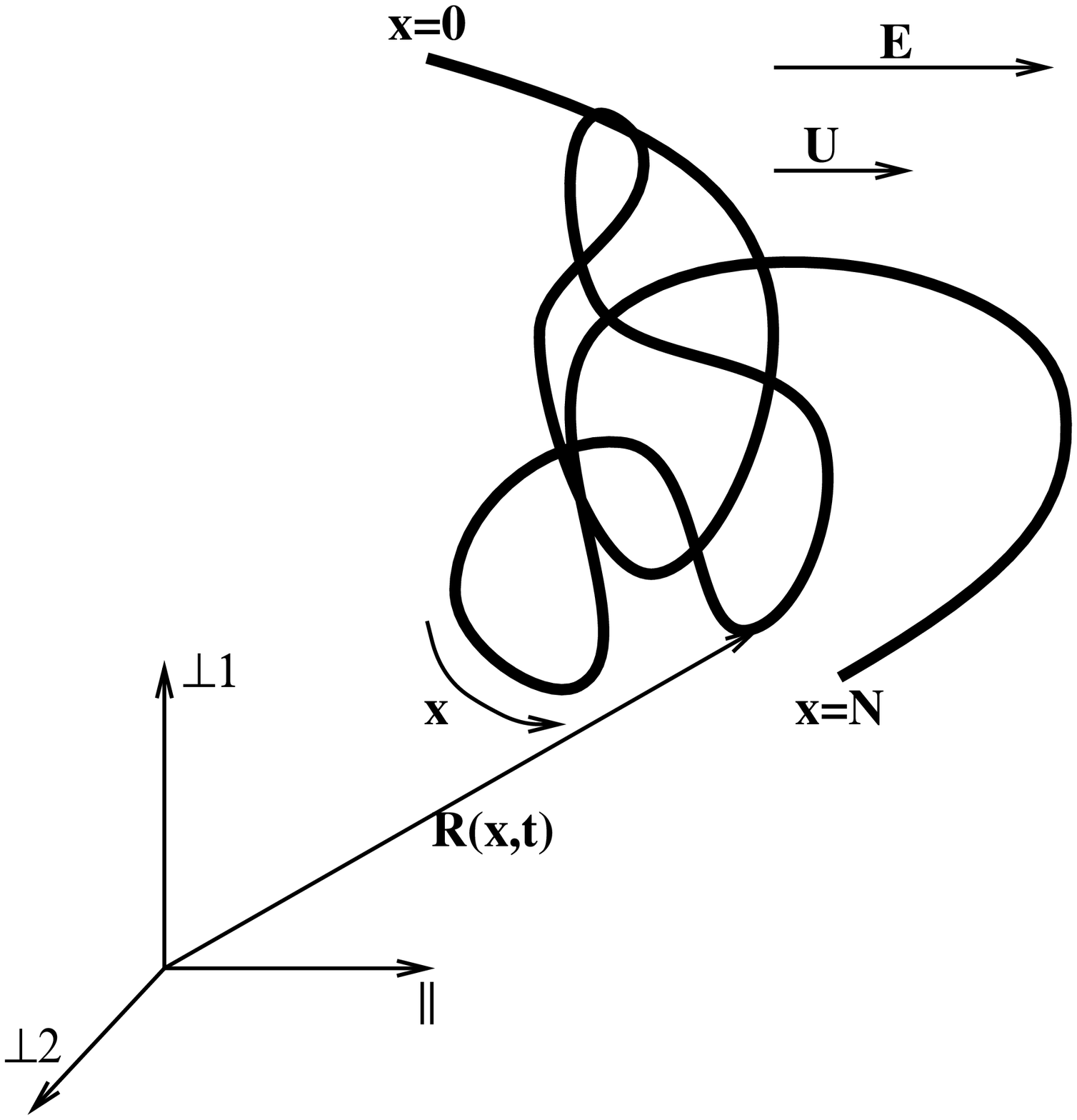}{The configuration of a polymer.}
One of the simplest is the
Rouse model\ref\rRs{P.E.~Rouse, J. Chem. Phys. {\bf 21}, 1272
(1953).}: The configuration of the polymer at
time $t$ is described by a vector ${\bf R}(x,t)$, where $x \in [0,N]$
is a continuous variable replacing the discrete monomer index (see Fig.\fPoly). 

Ignoring inertial effects, the relaxation of the polymer in a viscous
medium is approximated by
\eqn\Rouseeq{
\partial_t{\bf R}(x,t)=\mu{\bf F(R}(x,t){\bf )} =
K\partial_x^2{\bf R}(x,t)+{\eta}(x,t),
}
where $\mu$ is the mobility. The force {\bf F} has a
contribution
from interactions with near neighbors that are treated as
springs.
Steric and other interactions are ignored. The effect of the
medium
is represented by the random forces $\eta$ with zero
mean.
The Rouse model is a linear Langevin equation that is easily
solved.
It predicts that the mean square radius of gyration,
$R_g^2=\langle
|{\bf R}-\langle {\bf R}\rangle |^2\rangle $, is proportional
to the
polymer size $N$, and the largest relaxation times scale as the
fourth power of the wave number, (i.e., in dynamic light
scattering
experiments, the half width at half maximum of the scattering
amplitude scales as the fourth power of the scattering
wave vector
{\bf q}). These results can be summarized as $R_g\sim N^\nu $
and
$\Gamma({\bf q}) \sim q^z$, where $\nu$ and $z$ are called the
{\it swelling} and {\it dynamic} exponents, respectively\ref
\rFoot{We have changed the notation to confer with the 
traditions of polymer science. $\nu$ is
$\zeta$ and $z$ is $z/\zeta$ in terms of the notation used previously.}.
Thus, for the Rouse Model, $\nu=1/2$ and $z=4$.

The Rouse model ignores hydrodynamic interactions mediated by
the
fluid.  These effects were originally considered by Kirkwood
and
Risemann\ref\rRis{J.~Kirkwood and J.~Risemann, J. Chem. Phys.
{\bf 16}, 565 (1948).}\ and later on by Zimm\ref\rZm{B.H.~Zimm, 
J. Chem. Phys. {\bf 24}, 269 (1956).}. The basic
idea is that the motion of each monomer modifies the flow
field
at large distances. Consequently, each monomer experiences an
additional velocity
\eqn\eZimm{
\delta_H\partial_t{\bf R}(x,t) = {1\over{8\pi\eta_s}}\int
dx'{{{\bf F}(x')r_{xx'}^2+({\bf F}(x')\cdot{\bf r}_{xx'})
{\bf r}_{xx'}}\over {|{\bf r}_{xx'}|^3}} \approx \int dx'
{\gamma\over{|x-x'|^\nu}}\partial_x^2{\bf R},
}
where ${\bf r}_{xx'}={\bf R}(x)-{\bf R}(x')$ and the final
approximation is obtained by replacing the actual distance
between
two monomers by their average value.  The modified equation is
still linear in ${\bf R}$ and easily solved.  The main result
is the speeding up of the relaxation dynamics as the exponent
$z$ changes from 4 to 3.  Most experiments on polymer 
dynamics\ref\rAdam{See, for example, M.~Adam and M.~Delsanti,
Macromolecules {\bf 10}, 1229 (1977).}\
indeed measure exponents close to 3. Rouse
dynamics is still important in other circumstances, such as
diffusion of a polymer in a solid matrix, stress and
viscoelasticity in concentrated polymer solutions, and is also
applicable to relaxation times in Monte Carlo simulations.

Since both of these models are linear, the dynamics remains
invariant in the center of mass coordinates upon the
application
of a uniform external force. Hence the results for a drifting
polymer are identical to a stationary one.  This conclusion is
in fact not correct due to the hydrodynamic interactions.
For example, consider a rodlike
conformation of the polymer with monomer length $b_0$ where 
$\partial_xR_\alpha=b_0t_\alpha$ everywhere on the polymer, so that the
elastic (Rouse) force vanishes. If a uniform force ${\bf E}$ per 
monomer acts on this rod, the velocity of the rod can be solved using
Kirkwood Theory, and the result is\rDoi 
\eqn\erod{
{\bf v} = {(-\ln{\kappa}) \over 4\pi\eta_s b_0}{\bf E}\cdot
\left[{\bf I} + {\bf t}{\bf t}\right].}
In the above equation, $\eta_s$ is the solvent viscosity, ${\bf t}$
 is the unit tangent vector,  $\kappa= 2b/b_0N$ is the ratio 
 of the  width $b$  to the half length $b_0N/2$ of the polymer. 
A more detailed calculation
of the velocity in the more general case of an arbitrarily shaped
slender body by Khayat and Cox\ref\rKhayat{R.E.~Khayat and R.G.~Cox, 
J. Fluid. Mech. {\bf 209}, 435 (1989).}\ shows that
{\it nonlocal} contributions to the hydrodynamic force, which depend 
on the whole shape of the polymer rather than the local orientation,
are ${\cal O}(1/(\ln{\kappa})^2)$. Therefore, corrections to
Eq.\erod\ are small when $N \gg b/b_0$. 

Incorporating this tilt dependence of polymer mobility requires adding
terms nonlinear in the tilt, $\partial_x {\bf r}$, to a {\it local} equation of
motion. Since the overall force (or velocity) is the only vector breaking
the isotropy of the fluid, the structure of these nonlinear terms must be
identical to eq.\eGrow. Thus in terms of the fluctuations parallel and
perpendicular to the average drift, we again recover the equations,
\eqn\eKPZE{\left\{\eqalign{
\partial_tR_\parallel &= U_\parallel +
K_\parallel\partial_x^2R_\parallel +
{{\lambda_\parallel}\over 2}(\partial_xR_\parallel)^2 +
{{\lambda_\times}\over 2}\sum_{i=1}^2(\partial_xR_{\perp i})^2
+
\eta_\parallel(x,t), \cr
\partial_tR_{\perp i} &= K_\perp\partial_x^2R_{\perp i} +
\lambda_\perp\partial_xR_\parallel\partial_xR_{\perp i} +
\eta_{\perp i}(x,t), \cr}\right.
}
where $\{\perp\!i\}$ refers to the 2 transverse
coordinates
of the monomer positions.  The noise is assumed to be white
and
gaussian but need not be isotropic, i.e.
\eqn\eNoise{\left\{\eqalign{
\langle\eta_\parallel(x,t)\eta_\parallel(x',t')\rangle &=
2T_\parallel\delta(x-x')\delta(t-t'), \cr
\langle\eta_{\perp i}(x,t)\eta_{\perp j}(x',t')\rangle &=
2T_\perp\delta_{i,j}\delta(x-x')\delta(t-t'). \cr}\right.
}
At zero average velocity, the system becomes isotropic and the equations of
motion must
coincide with the Rouse model.  Therefore,
$\{\lambda_\parallel,
\lambda_\times,\lambda_\perp,U,K_\parallel-K_\perp,T_\parallel
-T_\perp\}$ are all proportional to $E$ for small forces.
The relevance of these nonlinear terms are determined by
the dimensionless scaling variable
$$ y=\left({U\over U^*}\right)N^{1/2}, $$
where $U^*$ is a characteristic microscopic velocity
associated with
monomer motion and is roughly 10-20 m/s for polystyrene in
benzene. The variable $y$ is proportional to another 
dimensionless parameter, the Reynolds number $Re$, which 
determines the breakdown of hydrodynamic equations and
onset of turbulence. However, typically $Re \ll y$, and the
hydrodynamic equations are valid for moderately large $y$.
Eqs. \eKPZE\ describe the static
and dynamical scaling properties of the nonlinear and
anisotropic
regime when $U>U^*N^{-1/2}$.

Eq.\eKPZE\ is just a slight variation from \eGro, with two
transverse components instead of one. Thus, the results 
discussed in the previous lecture apply. A more detailed
calculation of the nonlinear terms from hydrodynamics\ref
\rEKapp{See Appendices A and B of our longer paper\rEKpoly.}\ 
shows that all three nonlinearities are positive for small driving forces.
In this case, the asymptotic scaling exponents are isotropic,
with $\nu=1/2$ and $z=3$. However, the fixed points of the 
RG transformation are in general anisotropic, which implies 
a kinetically induced form birefringence {\it in the absence 
of external velocity gradients}. This is in marked contrast 
with standard theories of polymer dynamics where a uniform 
driving force has essentially no effect on the internal modes 
of the polymer.

When one of the nonlinearities approaches to zero, the swelling
exponents may become anisotropic and the polymer elongates or
compresses along the longitudinal direction.  However, the
experimental path in the parameter space as a function of $E$
is not known and not all of the different scaling regimes correspond
to actual physical situations.  The scaling results found by the
RG analysis are verified by direct integration of equations, as
mentioned in the earlier lectures. A more detailed discussion 
of the analysis and results can be found in our earlier
work\rEKpoly.

In constructing equations \eKPZE, we only allowed for local
effects, and ignored the nonlocalities that are the hallmark of
hydrodynamics. One consequence of hydrodynamic interactions 
is the {\it back-flow} velocity in Eq.\eZimm\ that can be added 
to the evolution equations \eKPZE. Dimensional analysis 
gives the recursion relation
\eqn\eZimmRR{\frac{\del\gamma}{\del\ell}=\gamma\left[
\nu z -1-(d-2)\nu\right]+O(\gamma^2),}
which implies that, at the nonlinear fixed point, this
additional term is surprisingly irrelevant for $d>3$, 
and $z=3$ due to the nonlinearities. For $d<3$, $z=d$ due to 
hydrodynamics, and the nonlinear terms are irrelevant. 
The situation in three dimensions is unclear,
but a change in the exponents is unlikely. Similarly, one 
could consider the effect of self-avoidance by including
the force generated by a softly repulsive contact potential 
\eqn\eContact{{b\over2}\int\,dx\,dx'\,{\cal V}
\left(\br(x)-\br(x')\right).}
The relevance of this term is also controlled by the scaling
dimension $y_b=\nu z -1-(d-2)\nu$, and therefore this effect is 
marginal in three dimensions at the nonlinear fixed point,
in contrast with both Rouse and Zimm models where
self-avoidance becomes relevant below four dimensions.
Unfortunately, one is ultimately forced to consider
non-local {\it and} nonlinear terms based on similar 
grounds, and such terms are indeed relevant below
four dimensions.
In some cases, local or global arclength conservation may be 
an important consideration in writing down a dynamics for the 
system. However, a 
local description is likely to be more correct in a more 
complicated system with screening effects (motion in a gel
that screens hydrodynamic interactions) where a first
principles approach becomes even more intractable. Therefore, 
this model is an important starting point towards understanding 
the scaling behavior of polymers under a uniform drift, a 
problem with great technological importance.
\medskip
\noindent{\bf Acknowledgments}
\smallskip
The work described here is part of the doctoral thesis of 
{\it Deniz Ertas} in the Physics Department of MIT.
Financial support from the NSF through grant number
DMR-93-03667 is gratefully acknowledged.
Many thanks are due to Drs. D. Kim, Y. Kim, J.M. Kim,
I.-M. Kim, H. Park, and B. Kahng for organizing the
4th CTP Workshop on Statistical Physics, and
providing the opportunity for these lectures.
\medskip

\bigskip\immediate\closeout\rfile                       
\noindent{\bf REFERENCES}\bigskip                                     
{\catcode`\@=11\escapechar=`  \input refs.tmp\vfill\eject} 
\end